\documentclass{aa}

\usepackage{graphicx, natbib}
\bibpunct{(}{)}{;}{a}{}{,} 
\usepackage[varg]{txfonts}
\usepackage[colorlinks=true,allcolors=blue]{hyperref}
\usepackage[normalem]{ulem}
\usepackage{threeparttable}

\newcommand{\edited}[1]{{#1}}

\begin{document}

\title{Burned to ashes:\\ How the thermal decomposition of refractory organics in the inner protoplanetary disc impacts the gas-phase C/O ratio}

   \author{Adrien Houge\inst{1}
    \and
      Anders Johansen\inst{1,2}
      \and
      Edwin Bergin\inst{3}
      \and
      Fred J. Ciesla\inst{4}
      \and
      Bertram Bitsch\inst{5}
      \and
      Michiel Lambrechts\inst{1, 2}
      \and
      Thomas Henning\inst{6}
      \and
      Giulia Perotti\inst{6, 7}
      }

    \institute{Center for Star and Planet Formation, GLOBE Institute, University of Copenhagen, Øster Voldgade 5-7, DK-1350 Copenhagen, Denmark\\
    \email{adrien.houge@sund.ku.dk}
    \and
    Lund Observatory, Department of Physics, Lund University, Box 43, 221 00 Lund, Sweden
    \and
    Department of Astronomy, University of Michigan, 1085 S. University, Ann Arbor, MI 48109, USA
    \and
    Department of the Geophysical Sciences, University of Chicago, Chicago, IL 60637, USA
    \and
    Department of Physics, University College Cork, Cork, Ireland
    \and
    Max-Planck-Institut f\"{u}r Astronomie (MPIA), K\"{o}nigstuhl 17, 69117 Heidelberg, Germany
    \and
    Niels Bohr Institute, University of Copenhagen, NBB BA2, Jagtvej 155A, 2200 Copenhagen, Denmark
    }
    
    \date{Received ; }

    \abstract
    {The largest reservoir of carbon in protoplanetary discs is stored in refractory organics, which thermally decompose into the gas-phase at the organics line, well interior to the water iceline. Because this region is so close to the host star, it is often assumed that the released gaseous material is rapidly accreted and plays little role in the evolution of the disc composition. However, laboratory experiments show that the thermal decomposition process is irreversible, breaking macromolecular refractory organics into simpler, volatile carbon-bearing compounds. As a result, unlike the iceline of other volatiles, which traps vapor inwards due to recondensation, the organics line remains permeable, allowing gaseous carbon to diffuse outward without returning to the solid phase. In this paper, we investigate how this process affects the disc composition, particularly the gas-phase C/H and C/O ratios, by incorporating it into a 1D evolution model for gas and solids, and assuming refractory organics dominantly decompose into C$_2$H$_2$. Our results show that this process allows this carbon-rich gas to survive well beyond the organics line (out to $7 \mathrm{~au}$ around a solar-mass star) and for much longer timescales, such that its abundance is increased by an order of magnitude. This has several implications in planet formation, notably by altering how the composition of solids and gas relate, and the fraction of heavy elements available to giant planets. In the framework of our model, refractory organics significantly influence the evolution of the gas-phase C/O ratio, which may help \edited{interpreting} measurements made \edited{with} Spitzer and JWST.}


    \titlerunning{Burned to ashes}
    \authorrunning{Houge et al.}
    
   \maketitle
   
\section{Introduction} 
\label{sec:intro}

The origin of life on Earth and other rocky planets is tightly related to the inward transport of carbon-bearing molecules through the protoplanetary disc \citep[e.g.,][]{oberg2021astrochemistry, krijt2023chemical}, and whether they are present in the rocky planet forming zone in solid or gas-phase \citep[][]{johansen2023anatomyIII}. In that context, the ratio of carbon-bearing to oxygen-bearing molecules, i.e., the C/O ratio, is an important quantity to measure, as it dictates the chemical pathway at play for both gas and solids in protoplanetary discs \citep[e.g.,][]{oberg2011effects, molliere2022interpreting}. Low C/O ratios yield molecules such as $\mathrm{H_2O}$ or $\mathrm{CO_2}$, while high C/O ratio favours the formation of hydrocarbon molecules such as $\mathrm{C_2H}$, $\mathrm{CH_4}$ and $\mathrm{C_2H_2}$ \citep[e.g.,][]{bergin2016hydrocarbon, kanwar2024hydrocarbon}. On top of that, the C/O ratio is fundamental in setting the thermal structure of planetary atmospheres \citep{madhusudhan2012CO}, along with the planet mineralogy and mantle dynamics, and hence ultimately the habitability \citep{unterborn2014role}.

Therefore, much effort has been dedicated to characterize observationally the gas-phase C/O ratio of the inner regions ($\lesssim 5 \mathrm{~au}$) of protoplanetary discs. Early results with Spitzer/IRS \citep[][]{werner2004spitzer} showed a trend where discs surrounding very low-mass stars ($<$$0.2 \mathrm{~M_\odot}$) display strong C$_2$H$_2$ emission and higher C/O ratio than discs around solar-type stars \citep[e.g.,][]{carr2008organics, pascucci2009different, salyk2011spitzer, pascucci2013atomic}. This trend seemed to be confirmed in the light of the first spectra obtained with JWST \citep[][]{tabone2023rich, arabhavi2024abundant, kanwar2024minds, long2024first}. However, \citet{colmenares2024jwst} recently demonstrated that high C/O can also be achieved in discs around solar-type stars, while discs around very low-mass stars and brown dwarfs can exhibit water-rich emission \citep{xie2023water, perotti2025browndwarf}. \edited{On top of that, a closer examination of sources previously characterised as rich in hydrocarbons but poor in water vapor showed that water is still present in abundance, though outshined in the spectra by hydrocarbon emissions \citep{arabhavi2025hiddenwater}.}

The cause of these variations in the C/O ratio is still poorly understood. \citet{walsh2015molecular} suggested that differences in the abundances of C$_2$H$_2$ and HCN can be related to the different FUV and X-ray-driven chemistry taking place around different stellar types. As such, variations in the C/O ratio would reflect the different partitioning of carbon in molecular species that may (or not) be observed, rather than variations in the total carbon content of the inner disc. \citet{najita2013hcn} also suggested that the formation of planetesimals outside the water iceline could lock oxygen-rich volatiles into planetesimals, decreasing their delivery to the inner regions hence increasing the inner disc C/O ratio \citep[see also][]{danti2023composition}.

Alternatively, it has been shown that the composition of protoplanetary discs is deeply influenced by the prolonged flux of inward-drifting icy pebbles, which carry a wealth of chemical species in volatile or refractory form \citep{booth2017chemical, krijt2018transport, booth2019planet, krijt2020co, schneider2021how, kalyaan2021linking, kalyaan2023effect, mah2023close, mah2024mind, williams2025COfuelled}. The largest fraction of the total carbon content is thought to be stored in solid dust grains \citep[e.g.,][]{mathis1977size, zubko2004interstellar, gail2017spatial}, mostly in the form of refractory organics, which are macromolecular species composed of C along with other elements such as H, N, O, S \citep[][]{bardyn2017carbon, krijt2023chemical} with an overall composition of approximately C$_{100}$H$_{75}$O$_{15}$N$_{4}$S$_{3}$ \citep[e.g.,][]{alexander2007origin, alexander2017nature, glavin2018origin}. In the inner part of protoplanetary discs, the temperature can become high enough \citep[$T_\mathrm{sub} = 300-500 \mathrm{~K}$,][]{nakano2003evaporation} for refractory organics to transition into the gas phase \edited{\citep{li2021earth, binkert2023carbon, penzlin2024bowie}, at a region often called "soot line" \citep[][]{kress2010soot}.} \edited{Apart from thermal processes, oxidation \citep[e.g.,][]{bauer1997simulation, finocchi1997chemical} and photolysis \citep{alata2014vacuum, anderson2017destruction} in the inner disc upper layers can also erode small dust grains and release carbon into the gas phase \citep{finocchi1997chemical, lee2010solar, anderson2017destruction, klarmann2018radial, binkert2023carbon, okamoto2024effects, vaikundaraman2025refractory}, though the efficiency of these processes is debated because only a small fraction of grains are lifted to sufficient heights to encounter atomic oxygen and UV radiations \citep{klarmann2018radial, binkert2023carbon}. In the end, thermal processes are the most reliable way to convert solid carbon into the gas-phase.} An obvious candidate to explain the abundance of gas-phase carbon-bearing species in the inner part of protoplanetary discs could thus be the thermal decomposition of refractory organics.

Recent studies attempted to explain the evolution of the C/O ratio in the inner part of discs using dust and volatile transport models, accounting for carbon stored in refractories or in volatiles, such as CH$_4$ \citep[e.g.,][]{mah2023close, mah2024mind, lienert2024changing}. However, \citet{mah2023close} found that the delivery of refractory carbon to the inner disc had little effect on the gas-phase C/O ratio, as it transitions into the gas phase only near the host star, which rapidly accretes the carbon-rich gas (see, however, next paragraph). Instead, they show that a high C/O ratio can be achieved around very low-mass stars thanks to the (smaller) fraction of the carbon content that is present in more volatile species, such as CH$_4$, which formed in the interstellar medium and was inherited by the protoplanetary disc \citep[][]{gibb2004interstellar}. As icy pebbles drift inward, they first release CH$_4$ in the outer regions, at the sublimation temperature of CH$_4$ around $30 \mathrm{~K}$ \citep[][]{lodders2003solar}, before shedding their water ice mantles at the water iceline in the inner disc. If the pebble flux eventually runs out, volatiles are no longer delivered and the oxygen-rich gas (mostly water vapor) accretes onto the host star, while the carbon-rich gas originating from the outer regions is still flowing inward. The accretion of water vapor occurs faster in discs around very low-mass stars, for which the viscous evolution is shorter and the water iceline lies closer-in.

However, these models treat the evolution of refractory organics similarly to that of any other chemical species. (1) When crossing their respective iceline, molecules sublimate off the surface of dust grains into the gas phase. (2) The plume of volatiles released inside the iceline creates a local overdensity, which diffuses in and out. (3) If vapor diffuses outwards and crosses back the iceline, it re-condenses on dust particles, locally increasing the solid surface density \citep[i.e., cold-finger effect,][]{stevenson1988rapid, cuzzi2004material, ros2013ice, drkazkowska2017planetesimal, bosman2022water, calahan2022water}. Nevertheless, refractory organics do not consist of a single molecular compound, but rather a diversity of complex carbon-rich species. Moreover, in the warm inner disc, refractory organics do not simply sublimate, intact, into the gas phase, but irreversibly decompose into simpler, more volatile molecules \citep[e.g., into $\mathrm{CH_4}$, $\mathrm{NH_3}$, see][]{kouchi2002rapid, nakano2003evaporation}. The irreversibility of the thermal decomposition process may strongly alter how refractory organics impact the evolution of the disc composition.

In this paper, we investigate how the irreversible thermal decomposition of refractory organics alters how these species contribute to the evolution of the disc composition, in particular the gas-phase C/O and C/H ratio. To do so, we use the 1D dust and volatile evolution model \texttt{chemcomp} \citep[][]{schneider2024chemcomp}, which accounts for dust coagulation and radial transport, along with the advection, diffusion, sublimation, and condensation of several molecular species. The code was adapted to simulate the evolution and thermal decomposition of refractory organics.

This paper is organised as follows. In Section \ref{sec:method}, we detail our dust and vapor evolution model based on \texttt{chemcomp}, along with our treatment of the thermal decomposition of refractory organics. We present our results in Section \ref{sec:results}, then discuss the implications of our findings in Section \ref{sec:discussion}, before giving our conclusions in Section \ref{sec:conclusion}. Appendix \ref{sec:appendix_disc_composition} gives further details on how atomic elements are partitioned into the different species evolved by \texttt{chemcomp}, while Appendix \ref{sec:appendix_pushing_param_space} explores further the parameter space of our disc model.

In the literature, it is common for the thermal processing of refractory organics to be referred to as "irreversible sublimation" \citep[e.g.,][]{li2021earth}, while the location where that happens is often called "soot line" \citep[e.g.,][]{kress2010soot}. Throughout this paper, we will use different definitions, describing the thermal processing of refractory organics as "thermal decomposition", and the location where that happens "(refractory) organics line". Though conceptually similar, we believe that our notations represent more accurately the physical processes at play.

\section{Model} 
\label{sec:method}

\subsection{Treatment of refractory organics}

As mentioned in Sect.~\ref{sec:intro}, dust particles carry the largest fraction of the disc carbon content in the form of refractory organics, macromolecular species mostly in the form of carbon in combination with other elements \citep[H, N, O, S,][]{alexander2017nature, bardyn2017carbon, glavin2018origin} which irreversibly decompose at high temperature \citep[$T_\mathrm{sub} = 300-500 \mathrm{~K}$,][]{nakano2003evaporation}. To compute the transport of refractory organics and the evolution of the disc composition, we employ the one-dimensional code \texttt{chemcomp} \citep{schneider2024chemcomp} that simulates the coagulation and radial transport of solids \citep[dust or pebbles, following the two-population algorithm by][]{birnstiel2012simple}, along with the radial transport (advection, diffusion), sublimation and condensation of several volatile species (e.g, $\mathrm{H_2O}$, $\mathrm{CH_4}$) in a 1D viscous $\alpha$-disc model \citep{shakura1973black, lynden1974evolution}.

We adapted \texttt{chemcomp} to account for the irreversible thermal decomposition of refractory organics, such that upon crossing the organics line, refractory organics are converted into gaseous $\mathrm{C_2H_2}$, a simpler, more volatile molecule characterised by $T_\mathrm{sub} = 70 \mathrm{~K}$ \citep[][]{penteado2017sensitivity}. For simplicity, we fix the decomposition temperature of refractory organics (hence the location of the organics line) at $T_\mathrm{sub} = 350 \mathrm{~K}$, within the range found by \citet{nakano2003evaporation}. However, for a complete treatment, the organics line should be treated as an organics band: a broad annulus spanning a range of temperatures ($T = 300-500 \mathrm{~K}$) in which organics made of different elemental composition decompose \citep[see figure 4 in][]{nakano2003evaporation}. We will focus on this in future works. It is currently unknown what would be the exact distribution of molecules after the thermal decomposition process. The choice of $\mathrm{C_2H_2}$ as a decomposition outcome is motivated by the fact that (1) refractory organics should be characterised by a high C/O ratio \citep[][]{nakano2003evaporation, alexander2017nature, gail2017spatial, li2021earth}, such that decomposition outcome should mostly be in the form of carbon-rich species, (2) $\mathrm{C_2H_2}$ is observed in abundance with JWST in the inner region of protoplanetary discs \citep[e.g.,][]{tabone2023rich, arabhavi2024abundant, kanwar2024minds, colmenares2024jwst}, and (3) thermo-chemical models find that carbon is often stored in $\mathrm{C_2H_2}$ molecules \citep{kanwar2024hydrocarbon, raul2025tracking}. We will discuss in Sect.~\ref{sec:discussion_other_recipient_andSc} the case where refractory organics decompose into CH$_4$ instead of C$_2$H$_2$. \edited{Note that we do not consider the processing of refractory carbon by any other processes than thermal decomposition, such as oxidation and photolysis \citep{finocchi1997chemical, alata2014vacuum}. We discuss this choice in Sect.~\ref{sec:discussion_alternative_destruction}.}

\subsection{Simulation setup}

We set up \texttt{chemcomp} over a discretised grid of $500$ radial cells log-spaced between either $0.01\mathrm{~au}$ or $0.1\mathrm{~au}$ to $1000~\mathrm{au}$.

Concerning the stellar and disc composition, we assume a solar composition following \citet{asplund2009chemical} \citep[see also e.g., Table 1 in][]{schneider2021how} such that the stellar C/O ratio is $0.55$, and a total metallicity $Z = 0.0142$ of all condensible species \citep{asplund2009chemical}. Atomic elements are partitioned into $18$ volatile and refractory species implemented in \texttt{chemcomp} (see more details in Appendix~\ref{sec:appendix_disc_composition}). We initially partitioned the total carbon reservoir among the main carbon-bearing species, as $29\%$ in CO, $10\%$ in $\mathrm{CO_2}$, $1\%$ in $\mathrm{CH_4}$ \citep[following measurements of interstellar ices and comets, see e.g.,][]{gibb2004interstellar, mumma2011chemical}, and $60\%$ in refractory organics\footnote{Though refractory organics are the dominant host of refractory carbon, carbon may still be found in other compounds, such as PAHs or amorphous carbon resisting to higher temperatures \citep[e.g.,][]{gail2017spatial}. We ran additional simulations with a different partitioning of carbon in Appendix~\ref{sec:appendix_carbon_partition}.} \citep{mathis1977size, zubko2004interstellar, bergin2015tracing, gail2017spatial, bardyn2017carbon}. Given that $\mathrm{C_2H_2}$ is only found as a trace species in the ISM \citep{ehrenfreund2000organic}, its initial abundance\footnote{Despite the initial abundance set to zero, $\mathrm{C_2H_2}$ can be found at the initial stage inside the organics line, as refractory organics present in this region decompose at the onset of the simulation.} is set to $0$, such that in the framework of our model it can only be produced by the thermal decomposition of refractory organics. Concerning the evolution of dust particles, we set the fragmentation velocity to $v_\mathrm{frag} = 1 \mathrm{~m~s^{-1}}$ in the whole disc \citep{gundlach2018tensile, musiolik2019contacts}.

We performed a set of simulations to investigate the impact of different parameters on our model. We explored the impact of the stellar mass $M_*$ (see Sect.~\ref{sec:results_COratio_param_space}), considering the two cases of either a solar-mass star or a low-mass star with a mass of $0.1 \mathrm{~M_\odot}$. Depending on the choice of the stellar mass, we adopted (i) the stellar luminosity $L_*$, following the stellar evolution models of \citet{baraffe2015new} at 1 Myr (leading respectively to $L_* = 1.933 \mathrm{~L_\odot}$ for the solar-mass star and $L_* = 0.067 \mathrm{~L_\odot}$ for the low-mass star), (ii) the initial disc mass $M_\mathrm{disc}$, to a constant fraction of the stellar mass $M_\mathrm{disc} = 0.1~ M_*$, and (iii) the disc characteristic radius $R_\mathrm{c}$, as $R_\mathrm{c} = 50 \mathrm{~au}$ for the solar-mass star and $R_\mathrm{c} = 25 \mathrm{~au}$ for the very-low mass star \citep{tobin2024observational}. Changes in the initial disc mass and disc characteristic radius notably modify the initial gas surface density $\Sigma_\mathrm{gas, 0}$, which follows \citep{lynden1974evolution}
\begin{equation}
\Sigma_\mathrm{g, 0} (r) = \dfrac{M_\mathrm{disc}(2-\gamma)}{2\pi R_\mathrm{c}^2} \bigg( \dfrac{r}{R_\mathrm{c}} \bigg)^{-\gamma} \mathrm{exp} \bigg[ -\bigg( \dfrac{r}{R_\mathrm{c}} \bigg)^{2-\gamma}\bigg],
\label{eq:surface_density_initial}
\end{equation}
where $\gamma \approx 1.08$ \citep{lodato2017protoplanetary}. We compute the disc midplane temperature with the stellar luminosity, and ignore the contribution of viscous heating given that it may be inefficient if disc winds dominate the angular momentum transport \citep{mori2019temperature}. We refer the reader to Appendix~\ref{sec:appendix_viscous_heating} for more \edited{discussion on viscous heating}.

We also explored the impact of the turbulence parameter $\alpha$, with values ranging from $\alpha = 10^{-4}$ to $\alpha = 10^{-2}$ (see Sect.~\ref{sec:results_COratio_param_space} and Appendix~\ref{sec:appendix_no_transfer_diff_alpha}). In the standard version of \texttt{chemcomp}, $\alpha$ governs the disc viscosity, turbulence velocity (hence particle size), and diffusion. In addition to the stellar mass and turbulence strength, we investigated further the parameter space of our model in Appendix~\ref{sec:appendix_pushing_param_space}, notably exploring different partitioning of carbon and a different fragmentation velocity for dust particles.

\edited{As we will see further in this manuscript, a consequence of our model is that the carbon-rich gas produced by the thermal decomposition of refractory organics can diffuse outwards, past the organics line. In that context, the Schmidt number Sc, defined as the ratio between advection and diffusion, is an important quantity to introduce, as its value directly affects the strength of outward transport by diffusion against the inward motion of gas by accretion. For high Schmidt number ($\mathrm{Sc \gtrsim 1}$), the inward advection of gas is stronger than diffusion and the outward transport of material operates on a limited radial scale. Previous studies attempted to quantify the Schmidt number with full 3D MHD simulations, finding values between $10$ \citep{carballido2005diffusion} and $0.85$ \citep{johansen2005dust}. However, \citet{pavlyuchenkov2007dust} argued that these simulations suffer from rather coarse numerical resolution, and that the Schmidt number could reach values around $\mathrm{Sc} = 1/3$. More recent 3D MHD simulations by \citet{zhu2015dust} that included non-ideal MHD effects also agree with values of the Schmidt number smaller than unity. In the standard implementation of \texttt{chemcomp} \citep{schneider2021how}, the different gaseous species (e.g., CO, H$_2$O) are treated separately, i.e., solving for the advection diffusion equation of individual components one by one. In that case, the Schmidt number is not a free parameter, and gaseous species evolve like it is fixed to $\mathrm{Sc} = 1/3$, in which case outward diffusion may operate on large distances. In this manuscript, we use the standard implementation of \texttt{chemcomp} with $\mathrm{Sc} = 1/3$. We refer the reader to Sect.~\ref{sec:discussion_varying_Sc} where we modified the numerical treatment of gas evolution in \texttt{chemcomp} to freely vary the Schmidt number. We test our new implementation in Appendix~\ref{sec:appendix_Schmidt_number}.}

\section{Results} 
\label{sec:results}
\subsection{Evolution of the main carbon and oxygen-bearing species}
\label{sec:results_SigmaD_SigmaG}

\begin{figure*}
    \centering
    \includegraphics[width=\textwidth]{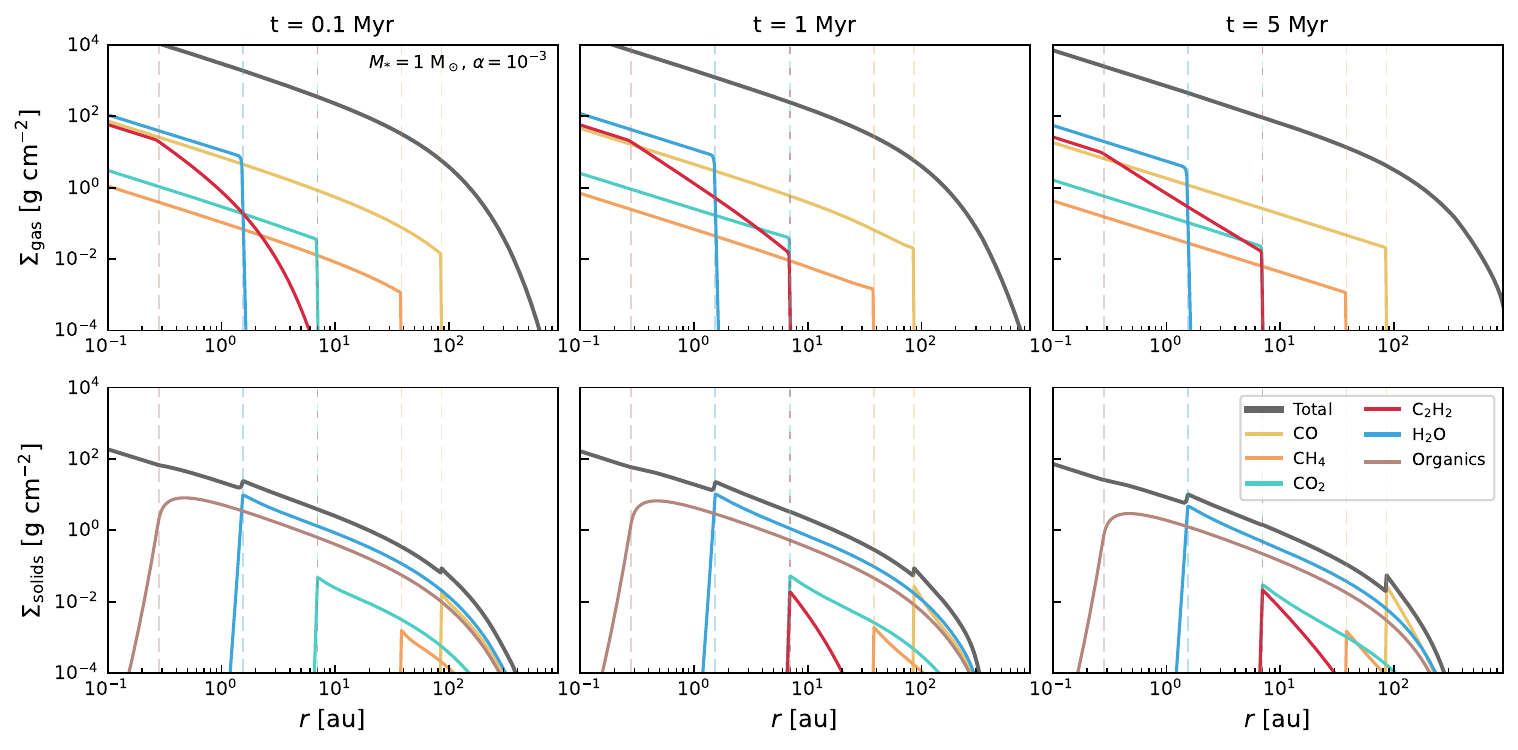}
    \caption{Gas surface density (top) and solid surface density (bottom) of the main chemical species that contribute to the C/O ratio at $0.1$, $1$, and $5 \mathrm{~Myr}$. We ran this simulation with $M_{*} = 1 \mathrm{~M_\odot}$ and $\alpha = 10^{-3}$. The total (dust or gas) surface density is represented as a dark solid line in each row. Vertical dashed lines indicate the sublimation lines of the displayed species in their respective color, using $T_\mathrm{sub, CO} = 20 \mathrm{~K}$, $T_\mathrm{sub, CH_4} = 30 \mathrm{~K}$, $T_\mathrm{sub, CO_2} = 70 \mathrm{~K}$, $T_\mathrm{sub, C_2H_2} = 70 \mathrm{~K}$, $T_\mathrm{sub, H_2O} = 150 \mathrm{~K}$, and $T_\mathrm{sub, refractory} = 350 \mathrm{~K}$ (see Table~\ref{tab:partition_abundances}).}
    \label{fig:SigmaD_SigmaG}
\end{figure*}

In Fig.~\ref{fig:SigmaD_SigmaG}, we present the gas surface density $\Sigma_\mathrm{gas}$ (upper panels) and solid surface density $\Sigma_\mathrm{solids}$ (lower panels) of the main species contributing to the C/O ratio, for a standard run where $\alpha = 10^{-3}$ and $M_* = 1 \mathrm{~M_\odot}$. When dust particles cross the organics line, refractory organics decompose into $\mathrm{C_2H_2}$, which cannot re-condense in the inner disc given its lower sublimation temperature \citep[i.e., $T_\mathrm{sub, C_2H_2}= 70 \mathrm{~K}$ similar to $\mathrm{CO_2}$,][]{penteado2017sensitivity}. In the absence of re-condensation right outside the organics line, the solid surface density of refractory organics does not display a peak at this location, unlike all other species. 

In the gas surface density plots (upper panels in Fig.~\ref{fig:SigmaD_SigmaG}), we see that the thermal decomposition of refractory organics (at $0.3 \mathrm{~au}$) results in a significant enhancement of gaseous $\mathrm{C_2H_2}$ in the inner disc, with abundances comparable to water vapor. The carbon-rich gas produced at the organics line diffuses inwards, but also outwards efficiently, up to the $\mathrm{C_2H_2}$ iceline located at $\sim$$7 \mathrm{~au}$. By $t=5$ Myr, the $\mathrm{C_2H_2}$ gas density profile has smeared out via diffusion. Outside its iceline, $\mathrm{C_2H_2}$ freezes out on the surface of dust grains, as can be seen in the lower panels in Fig.~\ref{fig:SigmaD_SigmaG}. In this model with $v_\mathrm{frag} = 1 \mathrm{~m~s^{-1}}$ and $\alpha = 10^{-3}$, the Stokes number of pebbles is rather low ($\mathrm{St_{frag}} \sim 0.001$ outside the $\mathrm{C_2H_2}$ iceline), resulting in an efficient outward diffusion of icy particles containing $\mathrm{C_2H_2}$. If the pebble size was larger (e.g., if $v_\mathrm{frag}$ > $1 \mathrm{~m~s^{-1}}$ or $\alpha \leq 10^{-3}$), particles containing $\mathrm{C_2H_2}$ ice would remain in a narrow region right outside its iceline.

\begin{figure}
    \centering
    \includegraphics[width=\columnwidth]{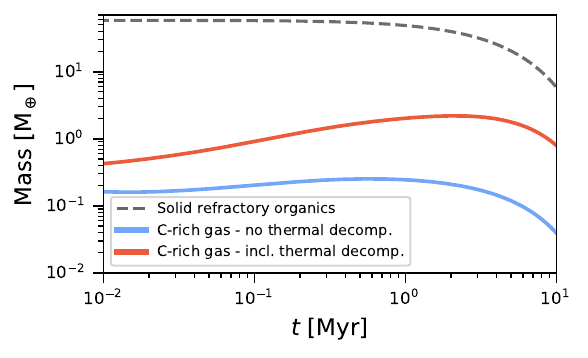}
    \caption{Evolution of the total mass of refractory organics and of the carbon-rich gas they release upon thermal processing at the organics line as a function of time in our fiducial disc setup ($\alpha = 10^{-3}$, $M_* = 1 \mathrm{~M_\odot}$). The blue line represents a standard model that excludes the thermal decomposition of refractory organics \citep[e.g.,][]{mah2023close}, where the carbon-rich gas released by refractory organics remains inside the organics line. Meanwhile, the red line shows the results from our model, including the thermal decomposition of refractory organics into gaseous C$_2$H$_2$. We see that the reservoir of carbon-rich gas originating from the thermal processing of refractory organics remains in the disc in higher abundance for a longer time when the irreversible decomposition process is included.}
    \label{fig:M_vs_time_tr_vs_notr}
\end{figure}

We show in Fig.~\ref{fig:M_vs_time_tr_vs_notr} the temporal evolution of the total mass of refractory organics and carbon-rich gas produced by their thermal processing in the disc, both for a standard model that does not include the irreversible thermal decomposition of refractory organics (blue lines) and for our model including that effect (red lines). The total mass of refractory organics decreases steadily with time. After 10 Myr, there still remains about $15\%$ of the initial reservoir of solids in the disc, as the Stokes number of pebbles is rather small ($\mathrm{St_{frag}} \sim 0.001$, see above), in which case the inward radial drift motion is slow. When the irreversible thermal decomposition is included, the carbon-rich gas (in the form of C$_2$H$_2$) created by the thermal processing of refractory organics is one order of magnitude more abundant than in models excluding that effect. This is due to outward diffusion, which takes material further out (up to $\sim$$7 \mathrm{~au}$ at the C$_2$H$_2$ iceline) where the timescale for viscous accretion $t_\mathrm{visc}$ is much longer. In fact, it can be written as \citep{shakura1973black}
\begin{equation}
\label{eq:viscous_timescale}
    t_\mathrm{visc} = \dfrac{r^2}{\nu},
\end{equation}
where $r$ is the radial distance to the star and $\nu = \alpha c_\mathrm{s}^2 \Omega_\mathrm{K}^{-1}$. While at the organics line ($0.3 \mathrm{~au}$), the viscous timescale approximates $0.06 \mathrm{~Myr}$, it is $1.5 \mathrm{~Myr}$ at the C$_2$H$_2$ iceline ($7 \mathrm{~au}$).

\subsection{Impact of refractory organics on the C/O ratio}

\subsubsection{Fiducial model}
\label{sec:results_COratio}

\begin{figure*}
    \centering
    \includegraphics[width=\textwidth]{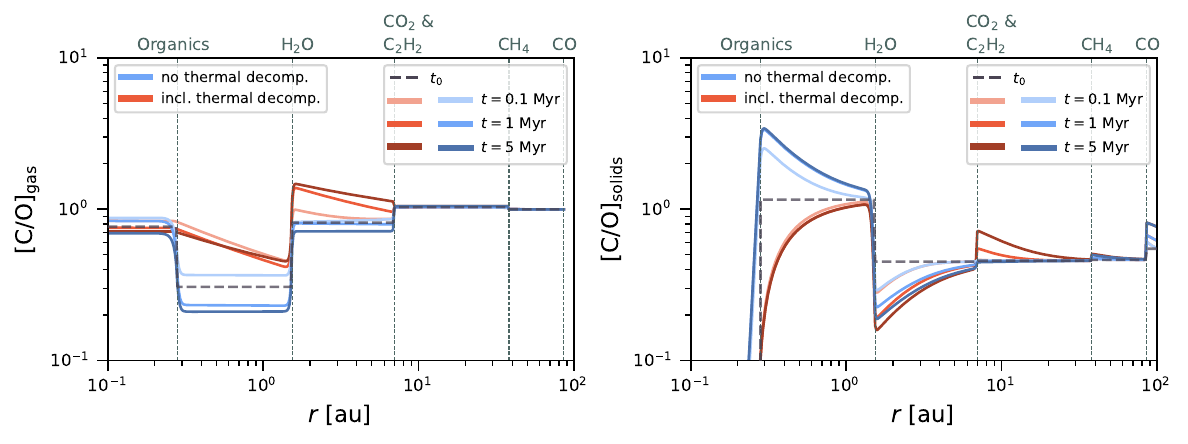}
    \caption{Gas-phase (left panel) and solid-phase (right panel) C/O ratios as functions of distance from a solar-mass star for a disc characterised by $\alpha = 10^{-3}$. Blue lines represent a standard model that excludes the thermal decomposition of refractory organics \citep[e.g.,][]{mah2023close}, while red lines show the results from our model, including that effect. The horizontal dashed grey line represents the initial C/O ratio. The sublimation lines of the main species contributing to the C/O ratio are indicated with vertical lines. From this plot, we see that the thermal decomposition of refractory organics significantly influences the gas-phase and solid-phase C/O ratio in the disc. The gas-phase C/O ratio is generally much higher when the thermal decomposition of refractory organics is included.}
    \label{fig:CO_vs_radius_tr_vs_notr}
\end{figure*}

We present in Fig.~\ref{fig:CO_vs_radius_tr_vs_notr} the evolution of the gas-phase and solid-phase C/O ratios as functions of distance to the star for our fiducial disc model ($M = 1 \mathrm{~M_\odot}$ and $\alpha = 10^{-3}$), both for a standard model that does not include the thermal decomposition of refractory organics (blue lines) and for our model including that effect (red lines).

We begin by focusing on the evolution of the gas-phase C/O ratio. In standard models excluding the thermal decomposition of refractory organics (blue lines in Fig.~\ref{fig:CO_vs_radius_tr_vs_notr}), the gas-phase C/O ratio is typically equal or slightly greater than unity outside the CO$_2$ iceline, where the main species in the gas-phase are carbon-bearing molecules like CO and CH$_4$. Meanwhile, it is smaller than unity inside the CO$_2$ iceline due to the sublimation of CO$_2$ and H$_2$O, both abundant oxygen-bearing species. Inside the organics line, the gas-phase C/O ratio rises again, typically to values slightly smaller than unity, as the mass fraction of refractory organics carried by pebbles is slightly lower than that of water (see Appendix~\ref{sec:appendix_disc_composition}).

In our model (red lines in Fig.~\ref{fig:CO_vs_radius_tr_vs_notr}), the thermal decomposition of refractory organics means that the organics line is no longer an impassable barrier to outwards transport. Carbon-rich gas from inside the organics line diffuses outwards out to the $\mathrm{C_2H_2}$ iceline, and increases the gas-phase C/O ratio up to this location. The overall gas-phase C/O ratio depends on the competition between the $\mathrm{C_2H_2}$ and the H$_2$O and CO$_2$ reservoirs. 

In between the organics line and water iceline, the gas-phase C/O ratio is much higher than in a standard model (typically greater than $\mathrm{[C/O]_{star}}=0.55$, hence super-stellar), but remains smaller than unity, due to the abundance of water vapor. As diffusion gradually smears out the $\mathrm{C_2H_2}$ gas density profile, the abundance of $\mathrm{C_2H_2}$ decreases with radial distance outside the organics line (see Fig.~\ref{fig:SigmaD_SigmaG}), leading to a slight decreasing slope towards the water iceline. Outside the water iceline, water freezes out on dust grains, leaving only CO$_2$ as an oxygen-bearing molecule in the gas-phase. CO$_2$ is less abundant than H$_2$O, and also contains carbon, such that it is less capable to lower the C/O ratio. It results in a sharp increase of the gas-phase C/O ratio above unity. We see a similar radial profile as inside the water iceline, with the gas-phase C/O ratio slightly decreasing with distance towards the CO$_2$ iceline. We also show in Appendix~\ref{sec:appendix_carbon_partition} the results from additional simulations with a different partitioning of carbon.

In standard models that exclude the thermal decomposition process (blue lines in Fig.~\ref{fig:CO_vs_radius_tr_vs_notr}), the solid-phase C/O ratio typically evolves anti-symmetrically to the gas-phase C/O ratio. In fact, the solid-phase C/O ratio gradually decreases up to the H$_2$O iceline, after which it sharply increases as there remains little oxygen-bearing species in the solid phase, while refractory organics still represent a large solid carbon reservoir. Inside the organics line, it drops as carbon can no longer be found in the solid phase. Outside icelines, the cold-finger effect (i.e., the re-condensation of gaseous molecules that diffuse back across their icelines) locally changes the solid-phase C/O ratio, including at the organics line. In our model (red lines in Fig.~\ref{fig:CO_vs_radius_tr_vs_notr}), the main differences are that the solid-phase C/O ratio does not peak right outside the organics line, as the cold-finger effect is inactive due to the irreversibility of the thermal decomposition process. We also notice that the solid-phase C/O ratio increases outside the C$_2$H$_2$ iceline, as the outward flow of C$_2$H$_2$ gas re-condenses on icy grains.

\begin{figure}
    \centering
    \includegraphics[width=\columnwidth]{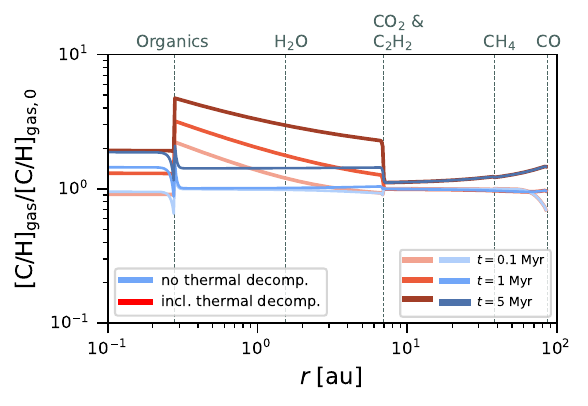}
    \caption{Gas-phase C/H ratio (normalised by the ratio at $t_0$) as a function of distance from a solar-mass star for a disc characterised by $\alpha = 10^{-3}$. Blue lines represent a standard model that excludes the thermal decomposition of refractory organics \citep[e.g.,][]{mah2023close}, while red lines show the results from our model, including that effect. The sublimation lines of the main species contributing to the C/O ratio are indicated with vertical lines. From this plot, we see that the thermal decomposition of refractory organics re-distributes gaseous carbon-bearing species through the disc, hence significantly influences the gas-phase C/H ratio.}
    \label{fig:CH_tr_vs_notr}
\end{figure}

We additionally represent in Fig.~\ref{fig:CH_tr_vs_notr} the evolution of the gas-phase C/H ratio (normalised by the initial value) with or without including the irreversible thermal decomposition of refractory organics. When the thermal decomposition process is included, C$_2$H$_2$-rich gas produced at the organics line diffuses outwards, up to the C$_2$H$_2$ iceline ($\sim$$7 \mathrm{~au}$). This increases the carbon content present in the gas phase by \edited{a factor $\sim$$2-5$}, compared to models excluding the thermal decomposition process. In the case where the decomposition outcome is a more volatile specie than C$_2$H$_2$ (e.g., CH$_4$, see Sect.~\ref{sec:discussion_other_recipient_andSc}), the gas-phase C/H can be increased on even greater distances.

\subsubsection{Exploring the parameter space}
\label{sec:results_COratio_param_space}

\begin{figure*}
    \centering
    \includegraphics[width=\textwidth]{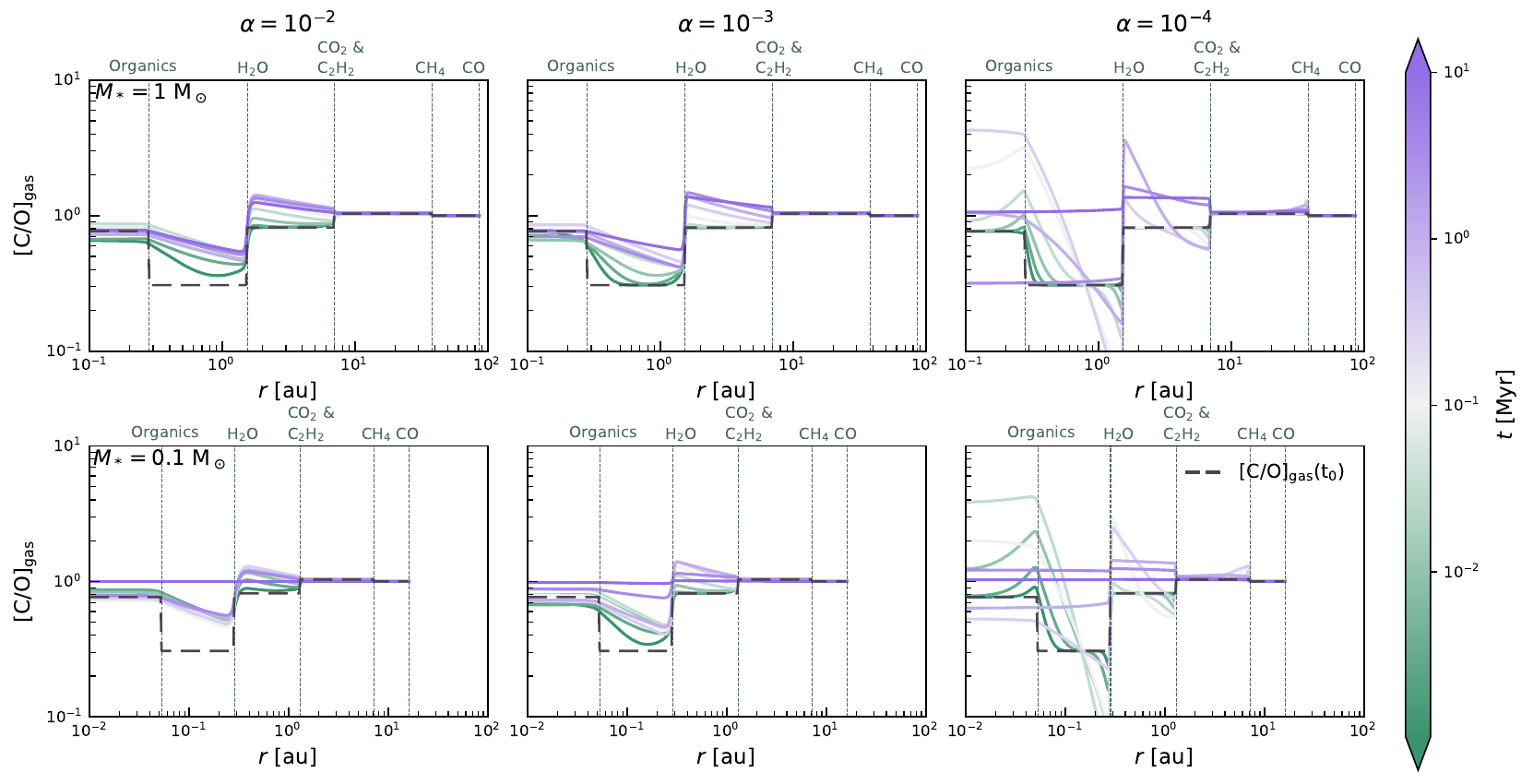}
    \caption{Gas-phase C/O ratio as a function of distance to the star for the two different values of the stellar mass ($M_* = 1,~0.1\mathrm{~M_\odot}$) and three values of the turbulent viscosity ($\alpha = 10^{-2},~10^{-3},~10^{-4}$). The horizontal dashed grey line represents the initial C/O ratio. The sublimation lines of the main species contributing to the C/O ratio are indicated with vertical lines. We see that the evolution of the gas-phase C/O ratio does not change significantly with varying disc turbulence ($\alpha > 10^{-3}$) and stellar mass. However, in the $\alpha = 10^{-4}$ case, where the pebble flux is intense and diffusion is slow, the variations in gas-phase C/O ratio are greater. Once the pebble flux runs out, viscous flows govern the evolution of the gas-phase C/O ratio, leading to differences depending on stellar mass.}
    \label{fig:CO_vs_radius}
\end{figure*}

\begin{figure*}
    \centering
    \includegraphics[width=\textwidth]{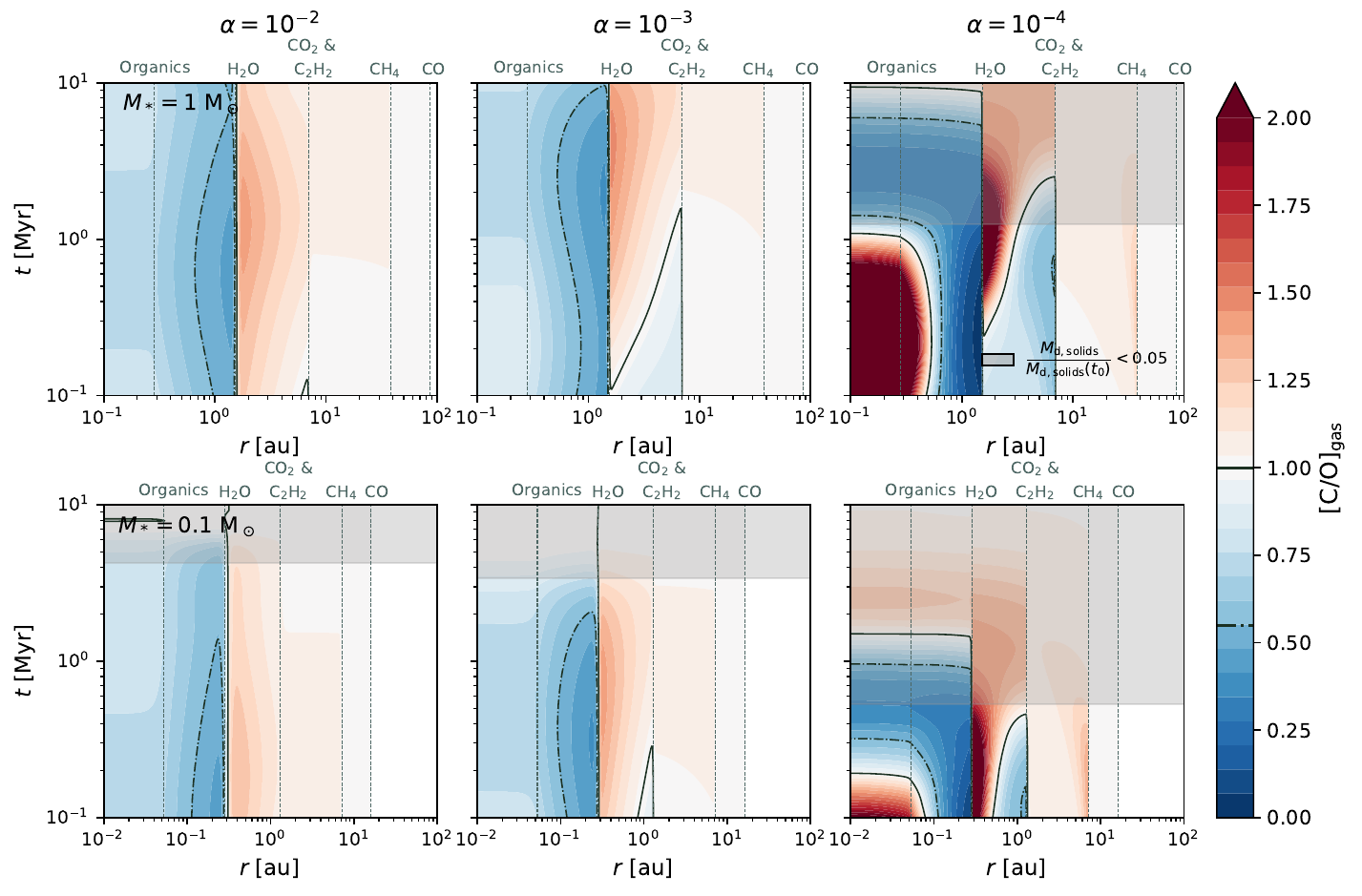}
    \caption{Gas-phase C/O ratio as a function of time and distance to the star for the two different values of the stellar mass ($M_* = 1,~0.1\mathrm{~M_\odot}$) and three values of the turbulent viscosity ($\alpha = 10^{-2},~10^{-3},~10^{-4}$). The sublimation lines of the main species contributing to the C/O ratio are indicated with vertical lines similarly to Fig.~\ref{fig:SigmaD_SigmaG}. The solid and dash-dotted contours indicate respectively $\mathrm{[C/O]}_\mathrm{gas}=1$ and the stellar value $\mathrm{[C/O]}_\mathrm{gas}=0.55$, above (resp. below) which the gas-phase C/O can be described as super-stellar (resp. sub-stellar). The shaded area indicates that the total mass of solids in the disc has decreased below $5\%$ of its initial value, marking the advanced depletion of the disc material. Similarly to Fig.~\ref{fig:CO_vs_radius}, we see that the gas-phase C/O ratio can be greater than unity (and often super-stellar) in different regions of the disc for most of the disc lifetime.}
    \label{fig:CO_vs_time_radius}
\end{figure*}

We present the evolution of the gas-phase C/O ratio as a function of distance to the star in Fig.~\ref{fig:CO_vs_radius} and as a function of time and distance in Fig.~\ref{fig:CO_vs_time_radius}, this time for different stellar mass and turbulence levels to investigate the impact of these parameters.

For all models, the radial profile of the gas-phase C/O ratio shows a similar behavior to what we discussed in Sect.~\ref{sec:results_COratio}, with a gas-phase C/O ratio slightly smaller than unity (and typically super-stellar) inside the water iceline, and greater than unity outside the water iceline. For strong ($\alpha = 10^{-2}$, left panels) and intermediate turbulence ($\alpha = 10^{-3}$, middle panels), it remains so for the entirety of the disc lifetime. Models with low turbulence ($\alpha = 10^{-4}$, right panels) show stronger variations, and even exhibit gas-phase C/O greater than unity inside the water iceline. This is because in low turbulence models: (1) the pebble size is larger, which intensifies the pebble flux hence the delivery of material, and (2) the advective and diffusive flows are weaker, such that delivered material is able to pile-up close to icelines.

Moreover, because the pebble size is larger, the pebble flux is also short-lived, such that the gas-phase C/O ratio evolves differently once the pebble reservoir runs out. This is estimated by the shaded area in Fig.~\ref{fig:CO_vs_time_radius}, which represents the time where the total disc mass of solids is smaller than to $5\%$ of its initial value, indicating significant depletion in the disc material. Beyond this time, a smooth protoplanetary disc would appear fainter and much more compact, i.e., making it complex to resolve or even detect with ALMA in the dust continuum. In this region, we enter a scenario similar to \citet{mah2023close}, where the carbon-rich gas inside the organics line is rapidly accreted once the pebble flux decreases, allowing water vapor to dominate and decrease the C/O ratio inside the water iceline for some time. However, in our case, the thermal decomposition process followed by the outward transport of carbon-rich gas outside the organics line allows it to survive for longer timescales in the disc (see also Fig.~\ref{fig:M_vs_time_tr_vs_notr}). As a result, the gas-phase C/O ratio is not as low as for models that do not include the thermal decomposition process (see also Appendix~\ref{sec:appendix_no_transfer_diff_alpha}). At the end of the disc lifetime, as found by \citet{mah2023close}, a high gas-phase C/O ratio in the inner disc can once again be achieved thanks to the inward flow of CO and CH$_4$-rich gas, though in our case, there is in addition a large reservoir of C$_2$H$_2$. As a result, the late-stage gas-phase C/O ratio is larger, and can reach values greater than unity within $10 \mathrm{~Myr}$ even in the solar-mass star model.

The stellar mass affects the timescale of the viscous evolution and the position of icelines, which move closer to the star around less luminous very-low mass stars (which is also why we lower the inner disc boundary to $0.01 \mathrm{~au}$ in these simulations). In our model, this does not significantly affect the evolution of the gas-phase C/O ratio, except for the low turbulence models. In that case, the pebble flux eventually runs out and the viscous evolution governs the evolution of the gas-phase C/O ratio. Because it is faster around low-mass stars, the time at which the water vapor reservoir disappears occurs earlier around lower-mass stars, i.e., at $t \sim 1 \mathrm{~Myr}$ for $M_* = 0.1 \mathrm{~M_\odot}$ instead of $t \sim 9 \mathrm{~Myr}$ for $M_* = 1 \mathrm{~M_\odot}$. 

Overall, the thermal decomposition of refractory organics into more volatile species has a great influence on the evolution of the C/O ratio in protoplanetary discs. It clearly depends on the details of gaseous transport (advection, diffusion), on the efficiency and lifetime of the pebble flux, and the mass fraction of refractory organics as compared to oxygen-bearing species. We will discuss the implications of our findings in Sect.~\ref{sec:discussion}, notably considering the observations of carbon-rich discs with JWST (Sect.~\ref{sec:discuss_JWST_CO}).

\section{Discussion} 
\label{sec:discussion}

\subsection{Gas-phase C/O ratio and the observed dichotomy with JWST}

\subsubsection{Evolution of the bulk C/O reservoir}
\label{sec:discuss_JWST_CO}

The gas-phase C/O ratio of the inner part of protoplanetary discs has been the subject of intense observational studies (Sect.\,\ref{sec:intro}). Early results with Spitzer/IRS, along with the first spectra obtained with JWST, found a trend with discs around very low-mass stars displaying higher C/O ratio than solar-type stars \citep[e.g.,][]{carr2008organics, pascucci2009different, salyk2011spitzer, pascucci2013atomic, tabone2023rich, perotti2023water, pontoppidan2024high, arabhavi2024abundant, long2024first, gasman2025minds}. \citet{mah2023close} provided a first theoretical model based on dust and volatile evolution to explain the trend between stellar mass and inner disc gas-phase C/O ratio. It relies on the fact that at the end of the lifetime of discs around low-mass stars (for which the viscous evolution is shorter, see Eq.~\ref{eq:viscous_timescale}), the oxygen-rich gas released by the sublimation of water ice in the inner disc has accreted onto the host star, while carbon-rich gas originating from the outer regions (CH$_4$ iceline at $r > 20-30 \mathrm{~au}$) continues to flow inwards. This causes the gas-phase C/O ratio in the inner disc of low-mass stars to increase in later stages, while solar-mass star preserves a low C/O ratio. 

However, to achieve high C/O ratio, the model of \citet{mah2023close} requires: (1) that pebbles in the outer regions are large enough to develop a high but short-lived pebble flux, such that the inflow of pebbles runs out rapidly, halting the delivery of oxygen to the water iceline, (2) to wait until the very end of the disc lifetime, when the water vapor reservoir has been accreted and carbon-rich gas from the outer disc flows inwards, and (3) an initial allocation of a larger fraction of the carbon content to CH$_4$ \citep[$\sim$$10\%$, though see also Appendix A.5 in][]{mah2023close}, greater than what is measured in interstellar ices and comets \citep[$\sim$$1\%$,][]{gibb2004interstellar, mumma2011chemical}. \edited{In addition, this model assumes that CH$_4$ remains in the gas phase for several Myr, while \citet{eistrup2016setting} showed that CH$_4$ gas can be converted into CO and CO$_2$ through chemical reactions already before reaching the inner disc.}

We have shown in this work that including the thermal decomposition of refractory organics in disc evolution models can significantly alter the evolution of the gas-phase C/O ratio in discs. In a similar disc setup to the one explored by \citet{mah2023close}, where the pebble size is large enough for the pebble flux to be intense and run out within the disc lifetime (e.g., for $v_\mathrm{frag} > 1 \mathrm{~m~s^{-1}}$ and/or $\alpha < 10^{-3}$), our results are similar inside the water iceline. After the pebble flux runs out, the disc viscous evolution becomes the main driver of changes in the gas-phase C/O ratio. The C/O ratio reaches values greater than unity in late stages, once the water vapor reservoir has been accreted. This occurs more rapidly in discs around very low-mass stars, as their viscous evolution is faster. The main difference with \cite{mah2023close} is that we also find a region with a high gas-phase C/O ratio outside the water iceline, where the C$_2$H$_2$ reservoir dominates over CO$_2$. In addition, the origin of the carbon-rich material is different. In their case, it comes from the fraction of carbon stored in CH$_4$, while in our case, the carbon-rich gas is sourced from the thermal processing of refractory organics in the inner disc. Those two reservoirs of carbon have rather different origins and histories, and we speculate whether this difference in the sourcing material may lead to different isotopic signature that may help disentangling the origin of hydrocarbons observed in the inner region of protoplanetary discs \citep{lee2025low}.

When pushing the exploration of the parameter space to intermediate and strong turbulence levels ($\alpha \geq 10^{-3}$), the pebble size is smaller and the pebble flux does not run out within the disc lifetime. In that case, the pebble flux remains the main driver of changes in the gas-phase C/O ratio and the stellar mass has little impact on its evolution. The overall gas-phase C/O ratio will thus depend on what is brought by pebbles, hence being dependent on how the reservoir of carbon initially stored in refractory organics compares to the reservoir of H$_2$O and CO$_2$. Inside the water iceline, water vapor remains in high abundance in the inner disc at all time, and maintains a gas-phase C/O ratio smaller than unity\footnote{If the water vapor reservoir was lowered (by a factor $\geq 2$, see Appendix \ref{sec:appendix_disc_composition}), e.g., due to the ice loss of dust grains by UV processing in the disc upper layers \citep{bergner2021ice, flores2025uv}, we could find C/O ratio greater than unity throughout the inner disc.}. Nevertheless, the thermal decomposition of refractory organics still \edited{facilitates} a higher carbon content in the gas-phase (Fig.~\ref{fig:CH_tr_vs_notr}) and \edited{to reaching} super-stellar gas-phase C/O ratio slightly inside the water iceline (see Fig.~\ref{fig:CO_vs_radius_tr_vs_notr} and Fig.~\ref{fig:CO_vs_time_radius}). Moreover, similarly to the low turbulence model, we are finding gas-phase C/O ratios greater than unity outside the water iceline for most of the disc lifetime, even for different turbulence levels and stellar masses. This reservoir could provide an explanation to the presence of a colder, more extended carbon-rich component observed in several discs with JWST \citep{arabhavi2024abundant, kanwar2024minds, colmenares2024jwst}. 

In our case, we do not find a strong dependency of the gas-phase C/O ratio on stellar mass, but rather on other parameters such as the \edited{probed region in the inner disc, the} turbulence strength and the partitioning of carbon and oxygen. This may go against the dichotomy observed with Spitzer and JWST \citep[e.g.,][]{carr2008organics, pascucci2009different, tabone2023rich}, though it is not unlikely that protoplanetary discs around different stellar masses could exhibit different physical and chemical conditions, e.g., if the disc turbulence correlates with stellar mass. Nonetheless, recent observational measurements have challenged the idea of a dichotomy exclusively based on stellar mass, with a solar-mass star displaying a high gas-phase C/O ratio \citep[DoAr 33, see][]{colmenares2024jwst}. And similarly, very low-mass stars, such as Sz 114 \citep[][]{xie2023water} and XX Cha \citep[see figure 5 in][based on unpublished JWST data]{long2024first}, are shown to have water-rich emission. Interestingly, DoAr 33 is characterised by a low accretion rate \citep[$\dot{M} \approx 10^{-10} \mathrm{~M_\odot~yr^{-1}}$,][]{cieza2010nature}, while Sz 114 and XX Cha have higher accretion rates \citep[$\dot{M} \approx 10^{-9} \mathrm{~M_\odot~yr^{-1}}$,][]{xie2023water, claes2022penellope}. This led \citet{colmenares2024jwst} to propose that the accretion rate is the fundamental parameter for the inner disc chemical content (i.e. C/O ratio) as opposed to stellar mass. 

\edited{Last, \citet{arabhavi2025hiddenwater} recently inspected further the spectra of discs around very low-mass stars, previously identified as hydrocarbon-rich but water-poor. They found that a significant water content is still present in those discs, but that it is outshined by hydrocarbon emission. The presence of water vapor indicates that pebble drift is still ongoing, in which case the simultaneous presence of hydrocarbons in high abundance cannot be explained by the inward flow of carbon-rich gas from the outer disc, which only occurs at the end of the disc lifetime. A more likely explanation is that the emission originates from inward-drifting pebbles, currently delivering both water and refractory organics to the inner disc. \citet{arabhavi2025hiddenwater} also suggested that the appearance of water-rich or hydrocarbon-rich protoplanetary discs may be related to the location of the water reservoir in the disc relative to the hydrocarbon reservoir. This is aligned with our findings (e.g., Fig.~\ref{fig:CO_vs_time_radius}), where both reservoirs co-exist and the outward diffusion of carbon-rich gas produces a region of high C/O ratio outside the water iceline.}

In the end, more work is required to make sense of the diversity in gas-phase C/O ratio observed in the inner part of protoplanetary discs, but it is clear that the thermal decomposition of refractory organics, the dominant host of carbon \citep[e.g.,][]{mathis1977size, zubko2004interstellar, gail2017spatial}, must significantly shape its evolution.

\subsubsection{Evolution of the observed C/O reservoir}
\label{sec:discuss_observed_CO}

Recently, \citet{houge2025smuggling} investigated the delivery of dust and volatiles to the inner part of protoplanetary discs due to pebble drift, focusing on the fraction of volatiles that may eventually be observed, i.e., in the surface layers above the dust optically thick region. They found that the observable water vapor reservoir can evolve differently from the bulk water vapor reservoir. Notably, in the presence of a traffic-jam effect inside the water iceline (e.g., due to a change in fragmentation velocity or higher dust densities), the observable water vapor reservoir remains almost constant in time, despite the intense ongoing delivery of vapor by icy pebbles. \citet{sellek2024co2} argued that the ratio of H$_2$O to CO$_2$ could be used as a proxy to infer the presence of a traffic-jam, as CO$_2$ is more volatile and its emission is less affected by the dust obscuration caused by the traffic-jam inside the water iceline.

In this work, we found that the C$_2$H$_2$ produced at the organics line by the thermal decomposition of refractory organics can diffuse outwards in the disc, past the water iceline, in regions less affected by the obscuration caused by a potential dust traffic-jam. Taking this effect into account, it is possible that the variations in gas-phase C/O ratio observed across a large sample of discs (see previous Section) could be caused by how carbon-bearing and oxygen-bearing molecules are distributed in the disc as compared to the dust, instead of representing actual changes in the bulk disc composition. Note, however, that the observation of C$_2$H$_2$ may be a signpost of bulk gas-phase C/O ratio greater than unity, as for C/O ratio smaller than unity, chemical networks tend to store carbon into species such as CO$_2$ \citep[e.g.,][]{bergin2016hydrocarbon, kanwar2024hydrocarbon}. In the end, caution is needed when interpreting observed gas-phase C/O ratio, and whether its variation across the disc population is due to obscuration effects or to actual differences in the bulk gas composition.



\subsection{Carbon abundance and planetary composition}
\label{sec:discuss_highCO_solid_gas}

An implication of the irreversible thermal decomposition of refractory organics
lies in how the composition of solids relates to that of the gas. It is usually assumed that the gas-phase composition is anti-symmetric to solids \citep{oberg2011effects}: e.g., if water vapor sublimates from the dust grains surface, the solid-phase C/O ratio is high, but the gas-phase C/O ratio is low. However, the outward diffusion of carbon-rich gas outside the organics line means that it is possible for both carbon-rich gas and carbon-rich solids to co-exist at the same location.

The re-distribution of carbon-rich gas also implies that the gas-phase C/H is increased in a larger fraction of the disc (see Fig.~\ref{fig:CH_tr_vs_notr}). Giant planets migrating inward while accreting surrounding gaseous material would therefore benefit from an extended window of opportunity to accrete carbon-rich gas. This will modify their heavy element content\footnote{The thermal processing of pebbles in the envelope of a pebble accreting protoplanet may also result in a carbon-rich atmosphere, without requiring carbon-rich gas to be present locally in the disc \citep[e.g.,][]{johansen2021pebble, wang2023atmospheric, johansen2023anatomyIII}.}, which may help explaining the enrichment of heavy elements observed in some giant exoplanets \citep[see also e.g.,][]{thorngren2016mass, schneider2021howII, bitsch2023enriching, danti2023composition}.

\subsection{What if refractory organics decompose into other compounds?}
\label{sec:discussion_other_recipient_andSc}

As discussed in Sect.~\ref{sec:method}, refractory organics thermally decompose around $T = 300 - 500 \mathrm{~K}$ in a variety of simpler and more volatile molecules \citep[][]{nakano2003evaporation}. For simplicity, we assume in this work that all the decomposed mass is converted into $\mathrm{C_2H_2}$. This choice is motivated by the observation in high abundance of this molecule with JWST in the inner part of protoplanetary discs \citep[e.g.,][]{tabone2023rich} and by thermo-chemical models showing that gas-phase chemistry rapidly stores carbon into $\mathrm{C_2H_2}$ \citep{kanwar2024hydrocarbon, raul2025tracking}. However, it is still likely that $\mathrm{C_2H_2}$ is subsequently converted to a range of organics with different sublimation temperatures as both CH$_3$ and C$_6$H$_6$ (i.e., benzene) are detected with JWST \citep{arabhavi2024abundant}. These two examples have disparate binding energies \citep[][$\sim$$2000$ K for CH$_3$ and $\sim$$4000$ K for benzene]{sameera2022modelling, clark2024hybrid} and therefore sublimation temperatures comparable to ammonia for benzene and closer to CO for CH$_3$ \citep{minnisale2022thermal}. Overall, it seems important to consider what would happen if carbon ended up in molecules with different sublimation temperatures.

\begin{figure}
    \centering
    \includegraphics[width=\columnwidth]{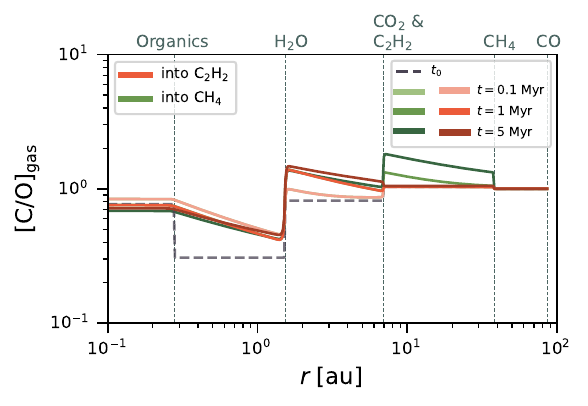}
    \caption{Gas-phase C/O ratio as a function of distance from a solar-mass star for a disc characterised by $\alpha = 10^{-3}$. Red lines represent our fiducial model where refractory organics thermally decompose into C$_2$H$_2$ ($T_\mathrm{sub} = 70 \mathrm{~K}$), while green lines show the results of a model where they decompose into CH$_4$, a more volatile specie ($T_\mathrm{sub} = 30 \mathrm{~K}$). The horizontal dashed grey line represents the initial C/O ratio. The sublimation lines of the main species contributing to the C/O ratio are indicated with vertical lines.}
    \label{fig:CO_vs_radius_ch4}
\end{figure}

We illustrate in Fig.~\ref{fig:CO_vs_radius_ch4} how the evolution of the gas-phase C/O ratio changes when refractory organics decompose into a more volatile specie, in this case chosen to be $\mathrm{CH_4}$ ($T_\mathrm{sub} = 30 \mathrm{~K}$). The sublimation temperature of $\mathrm{CH_4}$ is much lower than $\mathrm{C_2H_2}$, setting the methane iceline around $\sim$$40 \mathrm{~au}$ instead of $\sim$$7 \mathrm{~au}$ for the $\mathrm{C_2H_2}$ iceline. As can be seen in Fig.~\ref{fig:CO_vs_radius_ch4}, it implies that the carbon-rich gas diffuses much further out in the disc, modifying the gas-phase C/O ratio through most of the disc. In addition, it further increases the survival of the carbon-rich gas produced by the thermal processing of refractory organics, from $t_\mathrm{visc} \approx 1.5 \mathrm{~Myr}$ at $7 \mathrm{~au}$ to $8.5 \mathrm{~Myr}$ at $40 \mathrm{~au}$ (for $\alpha=10^{-3}$, see Eq.~\ref{eq:viscous_timescale}).

Besides carbon, refractory organics also contain, in small amount, atoms such as oxygen and nitrogen, such that they could also decompose into oxygen-bearing or nitrogen-bearing molecules. By assuming that refractory organics decompose only into C$_2$H$_2$, we neglected the contribution of their oxygen fraction to the C/O ratio. Nevertheless, this should not significantly affect our results concerning the evolution of the C/O ratio, as allocating oxygen to refractory organics would be counterbalanced by lower abundances of H$_2$O (see Appendix~\ref{sec:appendix_carbon_partition}). Concerning nitrogen, it is thought that a significant fraction of the nitrogen inventory may be stored in refractories \citep[][]{alexander2017nature, rice2018exploring, rubin2019elemental}, to abundances observable with JWST. Nitrogen-bearing molecules could be in the form of $\mathrm{NH_3}$ or $\mathrm{CN}$ \citep[][]{nakano2003evaporation, gail2017spatial}, and chemical reaction could quickly store nitrogen in secondary generation molecules such as HCN \citep[][]{wei2019effect}. Note that on top of the nitrogen fraction in refractories, more nitrogen may be stored in ammonium salts\footnote{Note that other mechanisms may irreversibly release volatile species in the inner part of protoplanetary discs, such as the trapping of CO$_2$ and CO \citep[][]{potapov2023formation, ligterink2024mind, bergner2024JWST} or noble gases \citep[][]{ciesla2018efficiency} within an H$_2$O ice matrix.} \citep[][]{altwegg2020evidence, poch2020ammonium, nakazawa2024nitrogen}.

\subsection{The influence of the Schmidt number on outward transport}
\label{sec:discussion_varying_Sc}

\begin{figure}
    \centering
    \includegraphics[width=\columnwidth]{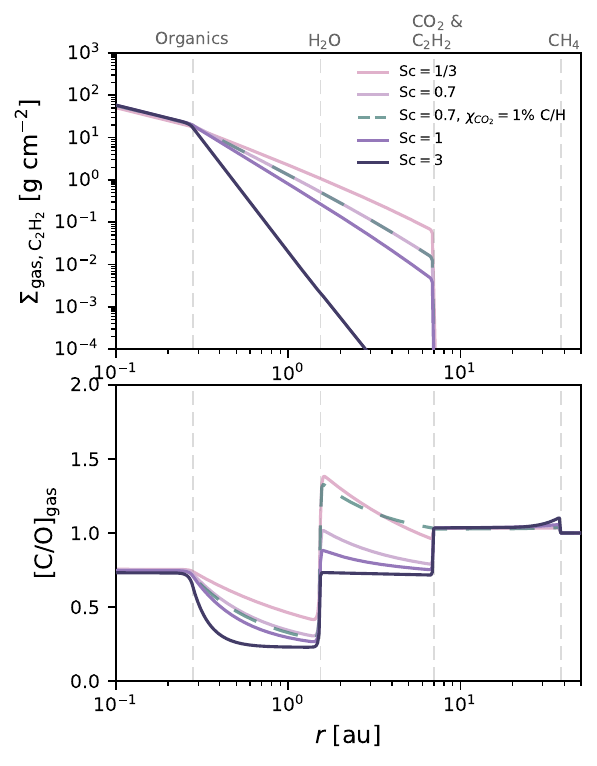}
    \caption{\edited{Gas surface density of C$_2$H$_2$ (top) and gas-phase C/O ratio (bottom) at 1 Myr for various Schmidt numbers for a disc around a solar-mass star with $\alpha = 10^{-3}$. We ran an additional model with a lower initial abundance of CO$_2$ and $\mathrm{Sc} = 0.7$ to show that it allows to obtain C/O ratios greater than unity outside the water iceline despite a higher Schmidt number. In that model, the mass fraction removed from CO$_2$ is given to CO, meaning that more oxygen becomes available for H$_2$O (see Table~\ref{tab:partition_abundances}). Inside the water iceline, the lowered abundance of CO$_2$ is compensated by the increased abundance of H$_2$O, such that the C/O ratio resembles the model with $\chi_\mathrm{CO_2} = 10\%$ and $\mathrm{Sc} = 0.7$. The sublimation lines of the main species contributing to the C/O ratio are indicated with vertical lines.}}
    \label{fig:SigmaG_CO_Sc}
\end{figure}

As discussed in Sect.~\ref{sec:method}, the strength of outward gas diffusion depends on the Schmidt number $\mathrm{Sc_{g}} = \nu/D_\mathrm{g}$, which is the ratio of the viscosity $\nu$ to the gas diffusivity $D_\mathrm{g}$. We used the standard implementation of \texttt{chemcomp} \citep[see more in][]{schneider2021how}, in which the gas surface density of all components $\Sigma_\mathrm{gas, Y}$ is evolved separately as
\begin{equation}
    \label{eq:gas_evo_std}
    \frac{\partial \Sigma_{\mathrm{gas, Y}}}{\partial t}-\frac{3}{r} \frac{\partial}{\partial r}\left[\sqrt{r} \frac{\partial}{\partial r}\left(\sqrt{r} \nu \Sigma_{\mathrm{gas, Y}}\right)\right]= \dot{\Sigma}_{\mathrm{Y}},
\end{equation}
\edited{where $\dot{\Sigma}_\mathrm{Y}$ encompasses the mass exchange between gas and solids due to sublimation and condensation. The total gas density $\Sigma_\mathrm{gas}$ is then obtained by summing the different components. This approach implies that the Schmidt number is not a free parameter and its value is fixed to $\mathrm{Sc_{g}} = 1/3$ \citep[see also][]{pavlyuchenkov2007dust}.} \edited{Though previous works argue in favor of a value of the Schmidt number smaller than unity \citep[e.g.,][]{johansen2005dust, pavlyuchenkov2007dust, zhu2015dust}, it could be higher than our fiducial $\mathrm{Sc_{g}} = 1/3$, reducing the strength of outward transport and altering how carbon-rich gas competes with other species to achieve C/O ratio greater than unity.}

\edited{To explore this aspect, we modified the implementation of gas evolution in \texttt{chemcomp}. As more than $\sim$$95\%$ of the gas disc mass is stored in the hydrogen and helium component, we consider this component to be the bulk gas background, which evolves with an equation like Eq.~\ref{eq:gas_evo_std}. Meanwhile, all other gaseous components evolved by \texttt{chemcomp} (see Table~\ref{tab:partition_abundances}) are now treated as tracers, advecting with the bulk gas and able to diffuse within it following}
\begin{equation}
\label{eq:gas_evo_Sc}
\frac{\partial \Sigma_{\mathrm{gas, Y}}}{\partial t} + \frac{1}{r} \frac{\partial}{\partial r} \left\{ r \Sigma_\mathrm{gas, Y}~{u}_{\mathrm{r, gas}} - r \Sigma_{\mathrm{gas}} D_\mathrm{g} \frac{\partial}{\partial r} \left( \frac{\Sigma_{\mathrm{gas, Y}}}{\Sigma_{\mathrm{gas}}} \right) \right\}
= \dot{\Sigma}_\mathrm{Y},
\end{equation}
\edited{where ${u}_\mathrm{r, gas}$ is the radial velocity of the bulk hydrogen and helium gas. We tested our new approach using $\mathrm{Sc_g} = 1/3$, and found that it reproduces well the gas evolution of the standard version of \texttt{chemcomp} (see Appendix~\ref{sec:appendix_Schmidt_number}).}

\edited{We present in Fig.~\ref{fig:SigmaG_CO_Sc} the gas surface density of C$_2$H$_2$ and the gas-phase C/O ratio at 1 Myr for different Schmidt numbers, for a disc around a solar-mass star with a turbulent viscosity $\alpha = 10^{-3}$.}

\edited{We see that the surface density of C$_2$H$_2$ outside the organics line decreases more sharply for increasing Schmidt number. For $\mathrm{Sc_g} = 0.7$, about twice the fiducial value, the amount of C$_2$H$_2$ reaching its iceline at $7 \mathrm{~au}$ is lowered by a factor $\sim$$5$. This makes it more challenging for the carbon-rich gas to compete with H$_2$O and CO$_2$, overall decreasing the gas-phase C/O ratio outside the organics line. In particular, outside the water iceline, the gas-phase C/O ratio becomes smaller than unity for $\mathrm{Sc} \gtrsim 0.7$. To still achieve C/O ratios greater than unity in this region, it is then required to lower the abundance of CO$_2$, e.g., through chemical reactions or by initially allocating a smaller fraction of the carbon reservoir to this molecule. In models presented in this manuscript, a fraction $\chi_\mathrm{CO_2} = 10\%$ of the carbon reservoir is given to CO$_2$. We see in Fig.~\ref{fig:SigmaG_CO_Sc} (dashed green line) that a smaller fraction $\chi_\mathrm{CO_2} = 1\% ~\mathrm{C/H}$ provides a gas-phase C/O ratio greater than unity even for $\mathrm{Sc} \gtrsim 0.7$. Exploring further the Schmidt number and alternatives to distribute carbon-rich gas outside the organics line when outward diffusion is inactive is clearly important and will be the subject of a follow-up work.}

\subsection{Thermal processing of refractory organics by FUor-type accretion outbursts}

Our model has focused on the thermal decomposition of refractory organics in the inner region of the protoplanetary disc, and has considered the organics line to be static ($\sim$$0.3 \mathrm{~au}$ around a solar-mass star). However, protoplanetary discs can experience large-scale thermal processing during FUor-type accretion outbursts \citep[][]{kenyon1990iras, dunham2012resolving, audard2014episodic, fischer2022accretion}. During such events, the mass accretion rate $\Dot{M}$ of the protostar increases by several orders of magnitude, which increases the disc temperature and pushes icelines outward, driving the thermal processing of ices and refractories on scales $\geq 10 \mathrm{~au}$ \citep[e.g.,][]{vanthoff2018methanol, lee2019ice, yamato2024chemistry, lee2024alma, calahan2024complex}. After the outburst, typically the volatiles, like H$_2$O, end up re-condensating on dust grains, unlike refractory organics for which thermal decomposition is irreversible. As a result, while particles can recover an oxygen-rich ice composition shortly after the outburst ended \citep{houge2023collisional, ros2024fast}, the gas-phase will remain filled with carbon-rich molecules that are unable to return into the solid phase. 

The recovery of the solid and gas composition may take long timescales, depending on how quickly unprocessed gas and dust grains from the outer regions move inward \citep[][]{colmenares2024thermal}. This, in turn, depends on how far the accretion outburst has processed material (i.e., the strength of the outburst). For an FUor-type accretion outburst of intermediate strength $L_* = 100 \mathrm{~L_\odot}$, the organics line is pushed to $r_\mathrm{orga, otb}$$~\gtrsim~$$2 \mathrm{~au}$. The drift timescale $t_\mathrm{drift}=r/v_\mathrm{r, peb}$ of carbon-rich un-processed pebbles from outside $r~$$>$ $r_\mathrm{orga, otb}$ is $230 \mathrm{~kyr}$ for $\alpha = 10^{-3}$, in which case the pebble Stokes number is $\mathrm{St_{frag}} \approx 0.001$. The viscous accretion timescale of the gas at $r_\mathrm{orga, otb}$ is larger, reaching $t_\mathrm{visc} \approx 400 \mathrm{~kyr}$ (see Eq.~\ref{eq:viscous_timescale}). In either case, the recovery of the solids and gas composition takes much longer than the typical outburst duration \citep[$\approx$$100 \mathrm{~yr}$]{audard2014episodic} and becomes comparable to the \edited{outburst rate \citep[$\approx 10^{-5} \mathrm{~yr^{-1}}$]{contreras2019determining}}, such that the disc composition up to $r_\mathrm{orga, otb}$ may never recover to initial conditions \citep[see also discussion on unrecovered annuli,][]{houge2023collisional}.

Meanwhile, the disc composition will be altered similarly to what we found in this work, with an increased gas-phase C/O ratio and a depletion of the reservoir of carbon in the solid phase, but on a more extended region than in the case of a static organics line located further in within a quiescent disc. The long-lasting changes in the gas-phase composition could be used as a tracer of past outburst in discs, similarly to other chemical tracers \citep[see e.g.,][]{jorgensen2013recent, molyarova2018chemical, wiebe2019luminosity, zwicky2024observational}. Moreover, as pointed by \citet{vanthoff2020carbon}, the depletion in carbon in solids in the inner au due to an outburst could explain the low carbon content of Earth by thermal processes \citep[see also][]{li2021earth}.

\subsection{Alternative destruction pathways of refractory carbon}
\label{sec:discussion_alternative_destruction}

In this paper, we have assumed refractory carbon to be stored in refractory organics, macromolecular species binding carbon with other atoms, such as nitrogen, which irreversibly transition into the gas-phase due to thermal decomposition around $300-500\mathrm{~K}$ \citep{nakano2003evaporation}. The presence of refractory organics in protoplanetary discs is inferred from their abundance in meteoritic and cometary material in the Solar System \citep{alexander2017nature, glavin2018origin}, but it remains under debate whether this material was inherited from the ISM \citep[e.g., due to the absence of C-N transitions in infrared spectra, see][]{juhasz2010dust, jang2024dust, henning2024minds, liu2025dust}. Alternatively, solid carbon atoms may be found directly within refractory carbon grains. In that case, refractory carbon can still transition into the gas-phase, but via other processes such as chemical sputtering with hydrogen at the surface of grains \citep[leading to e.g., gaseous C$_2$H and C$_2$H$_2$][]{lenzuni1995dust, borderies2025dust}, along with oxidation and photolysis \citep[e.g.,][]{finocchi1997chemical, lee2010solar, anderson2017destruction, klarmann2018radial, binkert2023carbon, okamoto2024effects, vaikundaraman2025refractory}. Nevertheless, like thermal decomposition, these processes are irreversible, and remain more efficient in the inner part of the protoplanetary disc, which is hotter and more exposed to the intense UV field from its host star. Overall, the processing of refractory carbon into the gas-phase by thermal decomposition or alternative processes (such as chemical sputtering) would yield similar results: carbon is transferred to simpler, more volatile carbon-bearing molecules in the gas-phase, able to diffuse outward in the protoplanetary disc and modify the C/H and C/O ratio on several astronomical units.

\section{Conclusions}
\label{sec:conclusion}

In this paper, we implemented the thermal decomposition of refractory organics, the dominant host of the total carbon reservoir, into a dust and volatile evolution model to investigate how this process impacts the protoplanetary disc composition, focusing on the C/H and C/O ratio given their significance in planet formation \citep[e.g.,][]{oberg2011effects}. Our findings are summarised as follows:

\begin{enumerate}
    \item The thermal decomposition of refractory organics into simpler, more volatile species allows for the outward re-distribution of the carbon-rich material past the organics line (Fig.~\ref{fig:SigmaD_SigmaG}), which significantly increases the carbon reservoir in the disc (Fig.~\ref{fig:CH_tr_vs_notr}).

    \item The timescale for gaseous material to viscously accrete onto the central star gets much longer with increasing radial distance. The outward diffusion of carbon-rich gas outside the organics line helps this material to survive in higher abundance on longer timescales in the disc, instead of being rapidly accreted by the central star (Fig.~\ref{fig:M_vs_time_tr_vs_notr}). 

    \item The thermal decomposition of refractory organics significantly impacts the evolution of the gas-phase C/O ratio (e.g., Fig.~\ref{fig:CO_vs_radius}). While the pebble flux is ongoing, the gas-phase C/O ratio remains smaller than unity inside the water iceline, but it is higher compared to models that do not include the thermal decomposition process (Fig.~\ref{fig:CO_vs_radius_tr_vs_notr}), reaching super-stellar values slightly inward of the water iceline (Fig.~\ref{fig:CO_vs_time_radius}). Outside the water iceline, the carbon-rich gas can overwhelm the CO$_2$ reservoir, leading to gas-phase C/O ratio greater than unity for most of the disc lifetime. As long as the pebble flux does not run out, the radial profile of the gas-phase C/O ratio is similar for a variety of turbulence levels and stellar masses.
    
    \item In the case of an intense but short-lived pebble flux (e.g., for $v_\mathrm{frag} > 1 \mathrm{~m~s^{-1}}$ and/or $\alpha < 10^{-3}$), our model aligns with the findings of \citet{mah2023close}, where the inner disc gas-phase C/O ratio reaches values greater than unity at late stages, and occurs faster around very-low mass stars. In that case, the main difference with \cite{mah2023close} is the origin of the carbon-rich material, as in our case it does not originate from an increased fraction of carbon in volatile CH$_4$, but is sourced from the thermal processing of refractory organics in the inner disc. We hypothesize whether this difference in the sourcing material may lead to different isotopic signatures that could be measured with observations (Sect.~\ref{sec:discuss_JWST_CO}).

    \item The significant impact of the thermal decomposition of refractory organics on the gas-phase C/O leads us to discuss if it could provide an explanation to observations made \edited{with} Spitzer and JWST \edited{indicating enhanced C/O ratios in some discs}. We find that the stellar mass does not have a strong effect on the inner disc gas-phase C/O ratio, \edited{which is rather determined} by other parameters like the turbulence strength, which would be in agreement with recent JWST results challenging the idea of a dichotomy purely based on stellar mass (see Sect.~\ref{sec:discuss_JWST_CO}). Additionally, the outward diffusion of C$_2$H$_2$ may explain the presence of a colder, more extended component of hydrocarbons observed in several discs \citep[e.g.,][]{colmenares2024jwst}. We also discussed whether variations in observed gas-phase C/O ratio may not be related to actual differences in the bulk disc composition, but in how carbon-rich and oxygen-rich gas are distributed compared to the dust population (see Sect.~\ref{sec:discuss_observed_CO}).

    \item The thermal decomposition of refractory organics allows for the simultaneous presence of carbon-rich solids and gas at a given \edited{radius}, and results in a large enhancement of carbon-rich gas in a wider region of the protoplanetary disc, which could impact the atmospheric properties (e.g., fraction of heavy elements) of planets forming in or migrating through these regions (see Sect.~\ref{sec:discuss_highCO_solid_gas}).

\end{enumerate}

Our findings illustrate the important role of refractory organics in the evolution of the disc composition. Future studies could implement a more complex distribution for the decomposition outcomes, for example including the release of NH$_3$ \citep[][]{nakano2003evaporation} and how that would impact the distribution of nitrogen in discs \citep[see also][for a recent work on ammonium salts]{nakazawa2024nitrogen}. Such studies would highly benefit from new laboratory experiments to help constraining the thermal decomposition of organics, to treat ultimately the organics line as an organics band: a broad annulus spanning a range of temperatures in which organics made of different elemental composition decompose \citep[see fig.4 in][]{nakano2003evaporation}. Also, \edited{the efficiency of outward diffusion is sensitive to the competition between viscosity and diffusivity (see Sect.~\ref{sec:discussion_varying_Sc}). A} more thorough exploration of the Schmidt number may help our understanding of outward diffusion. \edited{This} could also include the 2D (radial-vertical) structure of the gas flow, as outward diffusion may occur preferentially in the disc midplane \citep[][]{ciesla2009two}. \edited{Together, these directions open up promising avenues to better constrain the properties and impact of refractory organics on the evolution of the disc composition and forming protoplanets.}

\begin{acknowledgements} 

We are grateful to the anonymous reviewer for their insightful comments which helped improve the manuscript. A.H. thanks María José Colmenares for useful discussions regarding the gas-phase C/O ratio and JWST observations. A.H. also thanks Elishevah Van Kooten for useful discussions concerning the processing of refractory organics. A.J. acknowledges funding from the Danish National Research Foundation (DNRF Chair Grant DNRF159) and the Carlsberg Foundation (Semper Ardens: Advance grant FIRSTATMO). This project has made use of the public code \texttt{chemcomp} \citep[][]{schneider2024chemcomp}.

\end{acknowledgements}

\bibliography{paper.bib}{}
\bibliographystyle{aa}

\begin{appendix}
        
\section{Disc composition}
\label{sec:appendix_disc_composition}

We assume our host star and protoplanetary disc to follow solar abundances \citep[][]{asplund2009chemical}, such that the C/O ratio of the host star equals $0.55$. Atomic elements are partitioned into the $18$ volatile and refractory species considered in our model, as can be seen on Table~\ref{tab:partition_abundances} along with their respective sublimation temperature. With this partitioning, the initial mass fraction of water ice on dust particles is $\zeta_\mathrm{H_2O} = 23\%$, while it is $\zeta_\mathrm{orga} = 12\%$ for refractory organics. 

\begin{table*}
\centering
    \caption{Sublimation temperature and initial volume mixing ratio of the chemical species included in our model.}
    \label{tab:partition_abundances}
    \begin{threeparttable}
    \begin{tabular}{c c c}
    \hline\hline
    Species & $T_{\text{sub}}$ {[}K{]} & Initial volume mixing ratio \\ \hline 
    CO & 20  & 0.29 $\times$ C/H  \\
    N$_2$ & 20  & 0.5 $\times$ N/H  \\
    CH$_4$ & 30 & 0.01 $\times$ C/H  \\
    CO$_2$ & 70 & 0.1 $\times$ C/H  \\
    C$_2$H$_2$ & 70 & 0.0  \\
    H$_2$S & 150 & 0.1 $\times$ S/H  \\
    H$_2$O & 150 & O/H - (3 $\times$ MgSiO$_3$/H + 4 $\times$ Mg$_2$SiO$_4$/H + CO/H \\
    & & + 2 $\times$ CO$_2$/H + 3 $\times$ Fe$_2$O$_3$/H + VO/H \\ 
    & & + TiO/H + 3$\times$Al$_2$O$_3$ + 8$\times$NaAlSi$_3$O$_8$ + 8$\times$KAlSi$_3$O$_8$) \\
    Refractory organics & 350 & 0.6 $\times$ C/H \\
    FeS & 704 & 0.9 $\times$ S/H  \\
    NaAlSi$_3$O$_8$ & 958 & Na/H  \\
    KAlSi$_3$O$_8$ & 1006 & K/H  \\
    Mg$_2$SiO$_4$ & 1354 & Mg/H - (Si/H - 3$\times$K/H - 3$\times$Na/H) \\ 
    Fe$_2$O$_3$ & 1357 & 0.5 $\times$ (Fe/H - 0.9 $\times$ S/H) \\
    VO & 1423 & V/H  \\ 
    MgSiO$_3$ & 1500 & Mg/H - 2$\times$(Mg/H - (Si/H - 3$\times$K/H - 3$\times$Na/H)) \\ 
    Al$_2$O$_3$ & 1653 & 0.5$\times$(Al/H - (K/H + Na/H)) \\
    TiO & 2000 & Ti/H  \\  \hline
    \end{tabular}
    \begin{tablenotes}
        \item Volume mixing ratios of each species are based on the works of \citet{madhusudhan2014toward}, \citet{bitsch2020influence}, and \citet{schneider2021how,schneider2021howII}, and their condensation temperatures are from \citet{lodders2003solar}. We adapted the standard implementation of \texttt{chemcomp} to follow C$_2$H$_2$ and refractory organics instead of NH$_3$ and C. The sublimation temperature of C$_2$H$_2$ is taken from \citet{penteado2017sensitivity}, while the value for refractory species is taken from \citet{nakano2003evaporation} for refractory organics. Moreover, we did not consider the evolution of Fe$_3$O$_4$.  Solar elemental abundances \citep{asplund2009chemical} are assumed.
    \end{tablenotes}
    \end{threeparttable}
\end{table*}

\section{Impact of other parameters on the gas-phase C/O ratio}
\label{sec:appendix_pushing_param_space}

\subsection{More about turbulence strength}
\label{sec:appendix_no_transfer_diff_alpha}

We already investigated the impact of stronger or weaker turbulence in Sect.~\ref{sec:results_COratio_param_space}, though without discussing how it relates to previous models that did not include the thermal decomposition of refractory organics. In this section, we present additional figures to offer a more in depth comparison.

\subsubsection{Total mass and survival of carbon-rich gas}

\begin{figure*}
    \centering
    \includegraphics[width=\textwidth]{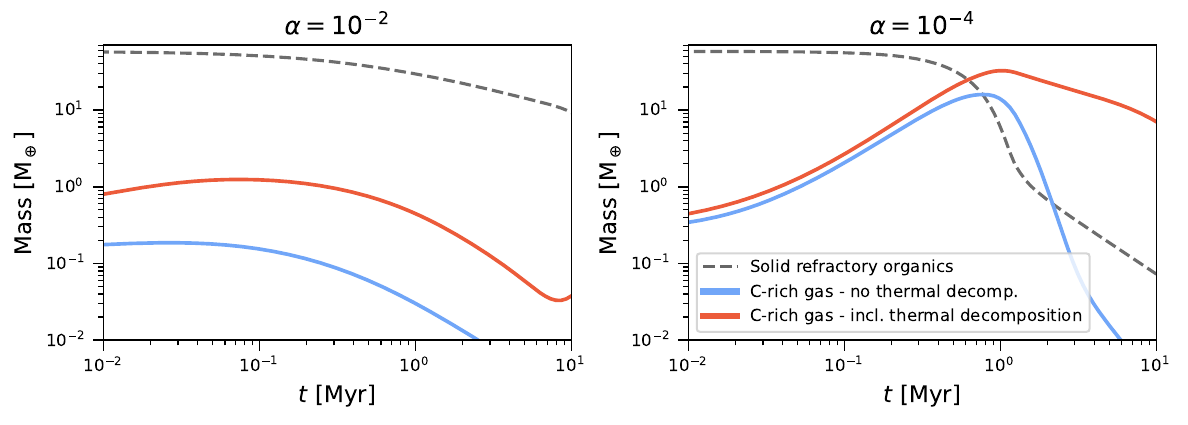}
    \caption{Evolution of the total mass of refractory organics and of the carbon-rich gas they release upon thermal processing at the organics line as a function of time in a disc around a solar-mass star characterised by $\alpha = 10^{-2}$ (left panel) or $\alpha = 10^{-4}$ (right panel). The blue line represents a standard model that excludes the thermal decomposition of refractory organics \citep[e.g.,][]{mah2023close}, where the carbon-rich gas released by refractory organics remains inside the organics line. Meanwhile, the red line shows the results from our model, including the thermal decomposition of refractory organics into gaseous C$_2$H$_2$.}
    \label{fig:mass_C2H2_vs_time_1e2_1e4}
\end{figure*}

We show in Fig.~\ref{fig:mass_C2H2_vs_time_1e2_1e4} the evolution of the total mass of refractory organics and carbon-rich gas (created by the thermal processing of refractory organics) with or without including the irreversible thermal decomposition process, similarly to Fig.~\ref{fig:M_vs_time_tr_vs_notr} but for strong ($\alpha = 10^{-2}$, left panel) and weak turbulence strength ($\alpha = 10^{-4}$, right panel).

As compared to the intermediate turbulence case, the strong turbulence model (left panel) shows similar results, with a rather slow decrease of the refractory organics due to the reduced particle size and pebble flux. When thermal decomposition is included, the carbon-rich gas (in the form of C$_2$H$_2$) created by the thermal processing of refractory organics is one order of magnitude more abundant than models excluding that effect, as outward diffusion takes material further out (up to $\sim$$7 \mathrm{~au}$ at the C$_2$H$_2$ iceline in our model) where the timescale for viscous accretion is much longer (see Eq.~\ref{eq:viscous_timescale}).

For the weak turbulence case, the reservoir of refractory organics sharply drops due to the intense pebble flux. The mass of carbon-rich gas is more than one order of magnitude greater than for intermediate and strong turbulence levels. After $\sim$$1 \mathrm{~Myr}$, the pebble flux has ran out, and the carbon-rich gas begins to the accrete. It is much slower when thermal decomposition is included, even more so in this model where weak turbulence is paired with slower viscous accretion ($t_\mathrm{visc} \propto \alpha^{-1}$, see Eq.~\ref{eq:viscous_timescale}). By $10 \mathrm{~Myr}$, there is $>4$ orders of magnitude more carbon-rich gas when thermal decomposition is included.

\subsubsection{Distribution of gaseous carbon}

\begin{figure*}
    \centering
    \includegraphics[width=\textwidth]{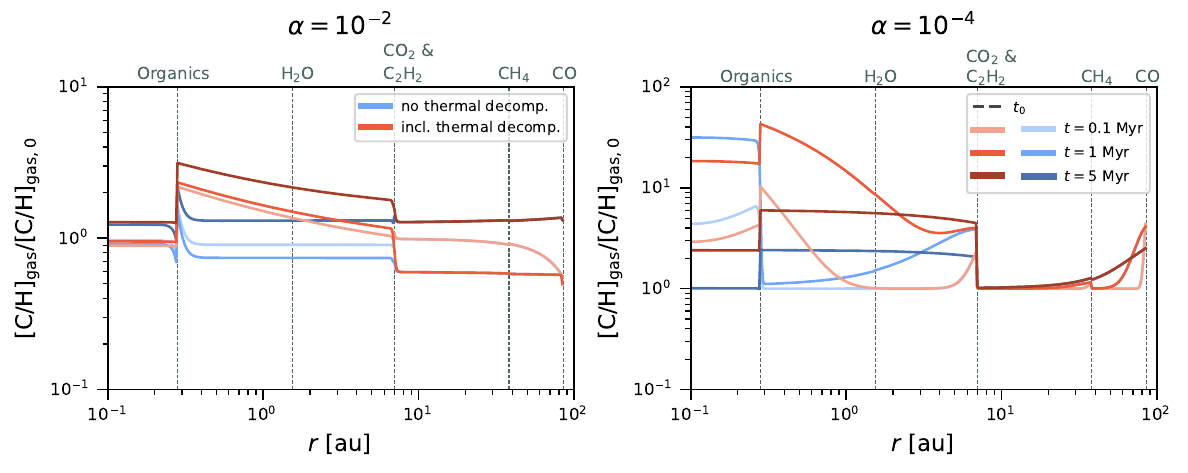}
    \caption{Gas-phase C/H ratio (normalised by the ratio at $t_0$) as a function of distance from a solar-mass star for a disc characterised by $\alpha = 10^{-2}$ (left panel) or $\alpha = 10^{-4}$ (right panel). Blue lines represent a standard model that excludes the thermal decomposition of refractory organics \citep[e.g.,][]{mah2023close}, while red lines show the results from our model, including that effect. The sublimation lines of the main species contributing to the C/O ratio are indicated with vertical lines.}
    \label{fig:CH_vs_time_radius_alpha1e2_4_notrsfr}
\end{figure*}

We present in Fig.~\ref{fig:CH_vs_time_radius_alpha1e2_4_notrsfr} the evolution of the gas-phase C/H ratio (normalised by its initial value) with or without including the irreversible thermal decomposition process, similarly to Fig.~\ref{fig:CH_tr_vs_notr} but for strong ($\alpha = 10^{-2}$, left panel) and weak turbulence strength ($\alpha = 10^{-4}$, right panel).

The high turbulence case shows similar results to the intermediate turbulence model (see Fig.~\ref{fig:CH_tr_vs_notr}, while the low turbulence case displays much greater enhancements in the gas-phase C/H ratio, more than an order of magnitude larger than models excluding the irreversible decomposition process.

\subsubsection{Carbon to oxygen ratio}

\begin{figure*}
    \centering
    \includegraphics[width=\textwidth]{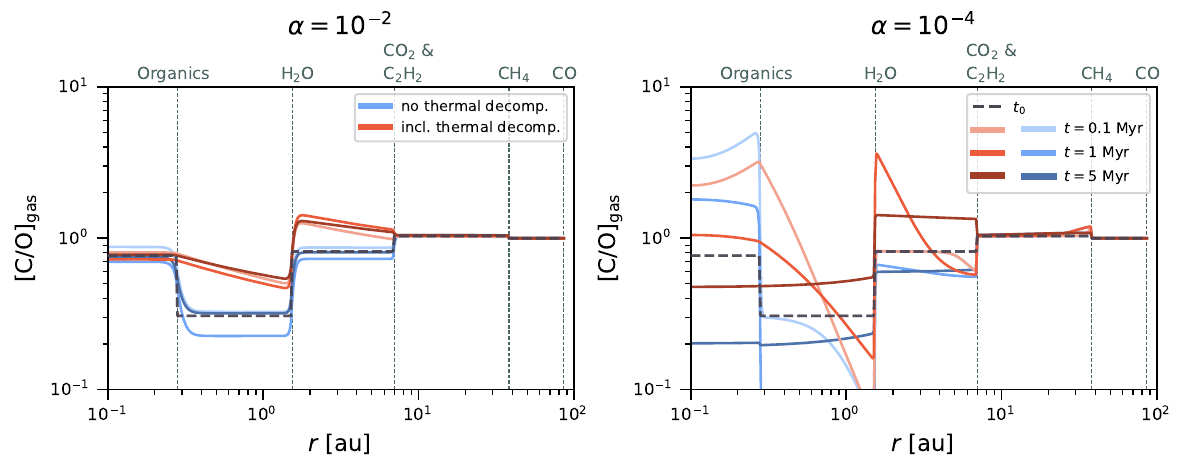}
    \caption{Gas-phase C/O ratio as a function of distance from a solar-mass star for a disc characterised by $\alpha = 10^{-2}$ (left panel) or $\alpha = 10^{-4}$ (right panel). Blue lines represent a standard model that excludes the thermal decomposition of refractory organics \citep[e.g.,][]{mah2023close}, while red lines show the results from our model. The horizontal dashed grey line represents the initial C/O ratio. The sublimation lines of the main species contributing to the C/O ratio are indicated with vertical lines.}
    \label{fig:CO_vs_radius_tr_vs_notr_turbulence24}
\end{figure*}

We present in Fig.~\ref{fig:CO_vs_radius_tr_vs_notr_turbulence24} the gas-phase C/O ratio, with or without including the irreversible sublimation of refractory organics, similarly to Fig.~\ref{fig:CO_vs_radius_tr_vs_notr}, but for stronger ($\alpha = 10^{-2}$, left panel) and weaker turbulence strength ($\alpha = 10^{-4}$, right panel).

Similarly to the intermediate turbulence case ($\alpha = 10^{-3}$, see Fig.~\ref{fig:CO_vs_radius_tr_vs_notr}), models that do not include the thermal decomposition of refractory organics are characterised by much lower gas-phase C/O ratio in between the organics line and C$_2$H$_2$ iceline than our models including that effect. As mentioned in Sect.~\ref{sec:results_COratio}, for weaker turbulence (right model), the pebble size is larger and the pebble flux high but short-lived. When the pebble reservoir runs out, the carbon-rich gas inside the organics line is rapidly accreted, and the gas-phase C/O ratio decreases. We also see that it decreases to much lower values when thermal decomposition is not included, as when it is included, the outward transport of carbon-rich gas allows it to survive for longer timescales in the disc (Fig.~\ref{fig:mass_C2H2_vs_time_1e2_1e4}), preventing it from rapid accretion (see 5 Myr lines in the right panel of Fig.~\ref{fig:CO_vs_radius_tr_vs_notr_turbulence24}) .

\subsection{Pebble fragmentation velocity}
\label{sec:appendix_frag_velocity}

\begin{figure}
    \centering
    \includegraphics[width=\columnwidth]{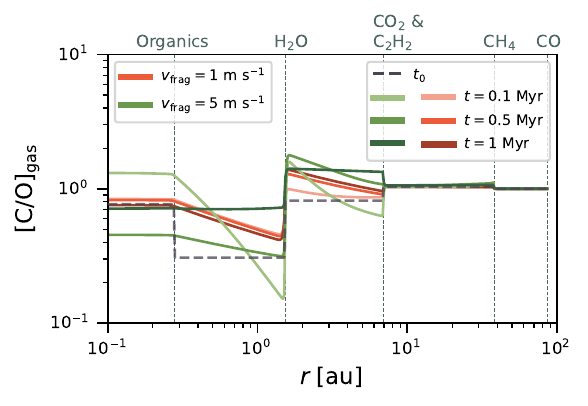}
    \caption{Gas-phase C/O ratio as a function of distance from a solar-mass star for a disc characterised by $\alpha = 10^{-3}$. Red lines represent our fiducial model where $v_\mathrm{frag} = 1 \mathrm{~m~s^{-1}}$, while green lines show the results of a model with $v_\mathrm{frag} = 5 \mathrm{~m~s^{-1}}$. The horizontal dashed grey line represents the initial C/O ratio. The sublimation lines of the main species contributing to the C/O ratio are indicated with vertical lines.}
    \label{fig:CO_vs_radius_vf1_5ms}
\end{figure}

We present in Fig.~\ref{fig:CO_vs_radius_vf1_5ms} the evolution of the gas-phase C/O ratio for our fiducial disc model ($M_* = 1 \mathrm{~M_\odot}$ and $\alpha = 10^{-3}$) and fiducial fragmentation velocity $v_\mathrm{frag} = 1 \mathrm{~m~s^{-1}}$ (red lines) compared to a model where dust particles are more resistant ($v_\mathrm{frag} = 5 \mathrm{~m~s^{-1}}$, green lines).

With a higher fragmentation velocity, the pebble size is larger \citep[$a_\mathrm{peb} \propto v_\mathrm{frag}^2$,][]{birnstiel2011dust}, leading to a strong but short-lived pebble flux. In that case, the variations in the gas-phase C/O ratio are more intense early on (see light red line in Fig.~\ref{fig:CO_vs_radius_vf1_5ms}) until the pebble reservoir runs out, much earlier in the disc lifetime ($t \sim 1 \mathrm{~Myr}$). These results are similar to what we found with weaker turbulence ($\alpha = 10^{-4}$) in Sect.~\ref{sec:results_COratio_param_space} \citep[$a_\mathrm{peb} \propto \alpha^{-1}$,][]{birnstiel2011dust}, though in that case the diffusive transport is also weaker resulting in an even stronger pile-up of carbon-rich gas.

\subsection{Viscous heating}
\label{sec:appendix_viscous_heating}

In the standard implementation of \texttt{chemcomp}, viscous heating can be included in the computation of the disc temperature. However, \texttt{chemcomp} computes the contribution of viscous heating at $t = 0$, and keeps its contribution fixed in time, despite variations in the disc properties, e.g., the gas density. On top of that, the contribution of viscous heating to the inner disc temperature may be inefficient if disc winds dominate the angular momentum transport \citep{mori2019temperature}. As a result, we decided not to include viscous heating in its current implementation, hence computing the disc temperature only considering stellar irradiation. 

\begin{figure}
    \centering
    \includegraphics[width=\columnwidth]{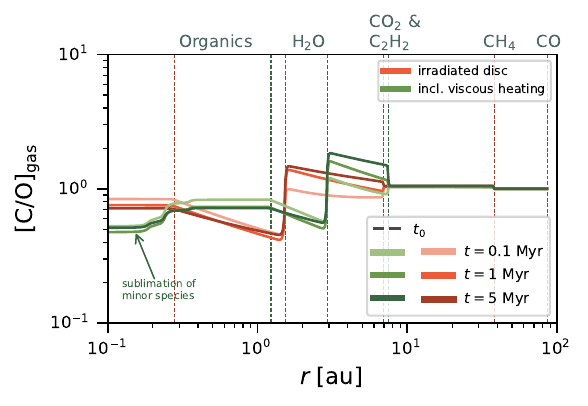}
    \caption{Gas-phase C/O ratio as a function of distance from a solar-mass star for a disc characterised by $\alpha = 10^{-3}$. Red lines represent our fiducial model where the disc temperature is only dependent on the stellar irradiation, while green lines show the results of a model that also includes viscous heating. The sublimation lines of the main species contributing to the C/O ratio are indicated with vertical dashed blue or red lines depending on the temperature model.}
    \label{fig:CO_vs_radius_with_without_viscous}
\end{figure}

We present in Fig.~\ref{fig:CO_vs_radius_with_without_viscous} the evolution of the gas-phase C/O ratio for our fiducial disc model ($M_* = 1 \mathrm{~M_\odot}$ and $\alpha = 10^{-3}$) for an irradiated disc (red lines, as used throughout this paper) and for a model that also includes the contribution of viscous heating (green lines). When viscous heating is active, it dominates the midplane temperature in the inner region ($\sim 2-3 \mathrm{~au}$), which mostly alters the location of the organics line and water iceline, pushing them further out. In the outer regions, viscous heating is much weaker, and the contribution from the stellar irradiation dominates, such that the icelines of e.g., C$_2$H$_2$ and CH$_4$ are located at similar distances from the host star in both models. Because the strength of viscous heating is dependent on the disc mass and turbulence, the transition where stellar irradiation comes to dominate over viscous heating would be located at smaller radii for disc models with weaker turbulences.

Concerning the evolution of the gas-phase C/O ratio, it is slightly higher in between the organics line and the C$_2$H$_2$ iceline when viscous heating is included. Inside $0.2 \mathrm{~au}$, the C/O ratio decreases as minor refractory oxygen-bearing species included in \texttt{chemcomp} begin to sublimate (see Table~\ref{tab:partition_abundances}).

\subsection{Distribution of carbon in carbon-bearing species}
\label{sec:appendix_carbon_partition}

As detailed in Sect.~\ref{sec:method}, we distribute the total carbon reservoir into the different carbon-bearing species implemented in \texttt{chemcomp}, with $29\%$ in CO, $10\%$ in $\mathrm{CO_2}$, $1\%$ in $\mathrm{CH_4}$ \citep{gibb2004interstellar, mumma2011chemical} and $60\%$ in refractory organics \citep{mathis1977size, zubko2004interstellar, bergin2015tracing, gail2017spatial, bardyn2017carbon}. As C$_2$H$_2$ is only found as a trace species in the ISM \citep{ehrenfreund2000organic}, its initial abundance is set to zero, and it can only be produced by the decomposition of refractory organics.

The partitioning of carbon is important because the fraction allocated to refractory organics can impact whether the C$_2$H$_2$ reservoir dominates over the H$_2$O and CO$_2$ reservoirs in the inner disc, hence increase the gas-phase C/O ratio above unity. If we attribute less carbon to refractory organics (e.g., $20\%$ less), it must be given to the other (volatile) carbon-bearing species: CO, CO$_2$, or CH$_4$. 

\begin{figure}
    \centering
    \includegraphics[width=\columnwidth]{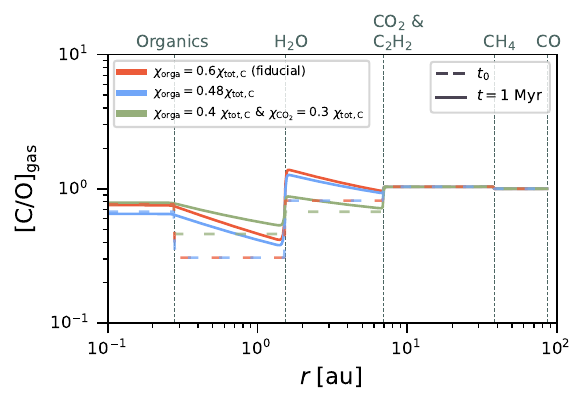}
    \caption{Gas-phase C/O ratio as a function of distance from a solar-mass star for a disc characterised by $\alpha = 10^{-3}$. Red lines represent our fiducial model where all the refractory carbon is into refractory organics ($\chi_\mathrm{orga} = 0.6 ~ \chi_\mathrm{tot, C}$, see Table~\ref{tab:partition_abundances}). Blue lines show the model where $20\%$ of the refractory carbon is given to a new compound ($T_\mathrm{sub} > 1000 \mathrm{~K}$) that does not sublimate within our disc boundary (such that $\chi_\mathrm{orga} = 0.48 ~ \chi_\mathrm{tot, C}$). Finally, green lines show a model where the fraction of carbon removed from refractory organics is given to CO$_2$. The sublimation lines of the main species contributing to the C/O ratio are indicated with vertical lines.}
    \label{fig:CO_vs_radius_less_refrac_C}
\end{figure}

If it is given to CH$_4$, we would find similar results to \citet{mah2023close}, such that a high gas-phase C/O ratio may be achieved at the end of the disc lifetime, once the pebble reservoir has emptied out. However, storing a too large fraction of the carbon content ($> 10\%$) in CH$_4$ would contradict measurements of interstellar ices and comets, finding that only a few percents of carbon is stored in CH$_4$ \citep{gibb2004interstellar, mumma2011chemical}. If it is given to CO or CO$_2$ instead, it would be paired with a greater sequestration of oxygen in these species instead of H$_2$O (see Table~\ref{tab:partition_abundances} to see how the initial water reservoir is determined), such that the lower abundance of refractory organics and C$_2$H$_2$ would be met with a similar decrease in the water abundance, overall not significantly affecting the gas-phase C/O ratio inside the water iceline. If it is given to CO$_2$ specifically, the gas-phase C/O ratio outside the water iceline would also be lowered as the CO$_2$ reservoir would more easily compete with the one of C$_2$H$_2$, overall flattening the radial profile of the gas-phase C/O ratio. We present the latter option in green shades in Fig.~\ref{fig:CO_vs_radius_less_refrac_C}.

At last, we considered the entirety of the refractory carbon to be stored as refractory organics, i.e., in the form of macromolecular material mainly composed of C along with other elements such as
H, N, O, S \citep{alexander2017nature, glavin2018origin}, with thermal decomposition temperatures around $300 - 500 \mathrm{~K}$ \citep{nakano2003evaporation}. Though refractory organics should, in fact, be the dominant reservoir of refractory carbon, a smaller fraction ($\sim 10-20\%$) may also be stored in hydrocarbons (such as PAHs) or amorphous carbon, that survive at much higher temperatures \citep[$>1000 \mathrm{~K}$][]{gail2001radial, wehrstedt2002radial}.

To consider that possibility, we ran an additional model where $80\%$ of the refractory carbon is stored in refractory organics, such that $\chi_\mathrm{orga} = 0.48 \chi_\mathrm{tot, C}$. The remaining is stored in a new carbon-bearing material with $T_\mathrm{sub} > 1000 \mathrm{~K}$ that does not sublimate within our disc model. We show the results of that model in blue shades in Fig.~\ref{fig:CO_vs_radius_less_refrac_C}. As the fraction of refractory organics (and C$_2$H$_2$ produced) is lowered without changing the fraction of carbon in other volatile species, the gas-phase C/O ratio overall decreases inside the C$_2$H$_2$ iceline by a factor comparable to the fraction removed from refractory organics.

\section{Testing our implementation of gas evolution}
\label{sec:appendix_Schmidt_number}

\begin{figure}
    \centering
    \includegraphics[width=0.9\columnwidth]{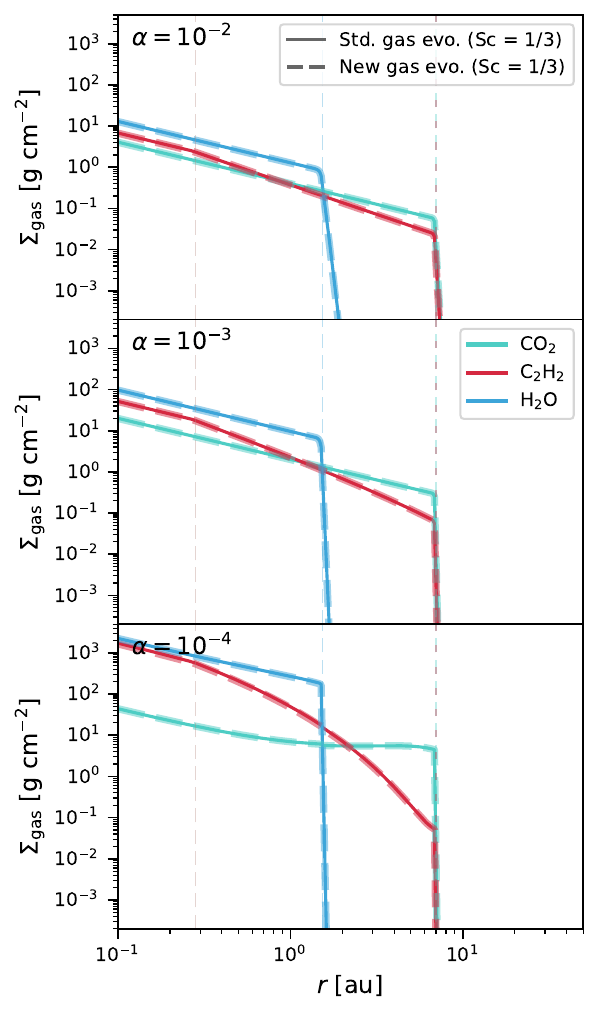}
    \caption{Gas surface density of H$_2$O, CO$_2$, and C$_2$H$_2$ at 1 Myr for a disc around a solar-mass star with varying turbulence viscosity levels. We compare results from the standard version of \texttt{chemcomp} (solid lines), where gaseous species are evolved separately, fixing the Schmidt number at $\mathrm{Sc_{g}} = 1/3$, to our new implementation (dashed lines) in which the Schmidt number is a free parameter, here also fixed to $\mathrm{Sc_{g}} = 1/3$. We see that the lines overlap, proving the validity of our new approach. Vertical dashed lines indicate the sublimation lines of the displayed species in their respective color, along with the organics line (dashed brown line).}
    \label{fig:SigmaG_testSc}
\end{figure}

\edited{We discussed in Sect.~\ref{sec:discussion_varying_Sc} that exploring the impact of the Schmidt number on outward transport required a modification of the gas evolution subroutine in \texttt{chemcomp}. The reason is that in the standard implementation of \texttt{chemcomp}, all gaseous components are evolved separately with Eq.~\ref{eq:gas_evo_std}, which implies that the Schmidt number is fixed to $\mathrm{Sc_{g}} = 1/3$. In our new implementation, the hydrogen and helium component of \texttt{chemcomp} is considered to be the bulk gas background, and evolves with Eq.~\ref{eq:gas_evo_std}. Meanwhile, all other gaseous components (e.g., CO, H$_2$O) are considered to be trace species advecting and diffusing within the bulk gas using Eq.~\ref{eq:gas_evo_Sc}. We present in Fig.~\ref{fig:SigmaG_testSc} the gas surface density of H$_2$O, CO$_2$, and C$_2$H$_2$ at 1 Myr for different turbulence levels for the two different gas evolution implementations. We fixed the Schmidt number to $\mathrm{Sc_{g}} = 1/3$ in our new implementation to compare with the standard version of \texttt{chemcomp}. We can see that the lines overlap for different species and turbulence strengths, proving the validity of our new approach.}

\end{appendix}

\end{document}